\documentclass[pra,twocolumn]{revtex4}
\usepackage{hyperref}
\usepackage{amssymb, amsmath}
\usepackage{graphicx}
\usepackage{color}
\usepackage{empheq}
\usepackage{lipsum} 
\usepackage{enumitem}
\setlist{nosep}

\newcommand{\avg}[1]{\left\langle {#1} \right\rangle}

\newcommand{\beq}{\begin{equation}}
\newcommand{\eeq}{\end{equation}}
\newcommand{\bea}{\begin{eqnarray}}
\newcommand{\eea}{\end{eqnarray}}

\newcommand{\comment}[1]{}

\newcommand{\binA}{\beta_{{\rm in}}}
\newcommand{\boutA}{\beta_{{\rm out}}}
\newcommand{\bin}[1]{\beta_{{\rm in},{#1}}}
\newcommand{\bout}[1]{\beta_{{\rm out},{#1}}}
\newcommand{\bbinA}{\bar{\beta}_{{\rm in}}}
\newcommand{\bboutA}{\bar{\beta}_{{\rm out}}}

\newcommand{\Amat}{\bar{A}}

\begin{document}

\title{Optical Devices based on Limit Cycles and Amplification in Semiconductor Optical Cavities}

\author{Ryan Hamerly}\email{rhamerly@stanford.edu}
\author{Hideo Mabuchi}
\affiliation{Edward L.\ Ginzton Laboratory, Stanford University, Stanford, CA 94305}
	
\date{\today}
	
\begin{abstract}
	At strong pump powers, a semiconductor optical cavity passes through a Hopf bifurcation and undergoes self-oscillation.  We simulate this device using semiclassical Langevin equations and assess the effect of quantum fluctuations on the dynamics.  Below threshold, the cavity acts as a phase-insensitive linear amplifier, with noise $\sim 5\times$ larger than the Caves bound.  Above threshold, the limit cycle acts as an analog memory, and the phase diffusion is $\sim 10\times$ larger than the bound set by the standard quantum limit.  
	We also simulate entrainment of this oscillator and propose an optical Ising machine and classical CNOT gate based on the effect.
\end{abstract}

\maketitle	

Many problems in simulation, optimization and machine learning are analog in nature and mapping them onto a digital processor incurs significant overhead.  As a result, there has been a recent revival of interest in analog or ``neuromorphic'' computing systems \cite{Utsunomiya2011, Tezak2014}.  Devices that can spontaneously oscillate are a key component in this neuromorphic architecture.  Such devices can function as an analog memory \cite{Tezak2014}, a phase-insensitive amplifier \cite{Kwon2013, KwonThesis}, or a complex-valued neuron \cite{HiroseBook}, among other things.  In addition, large networks of such oscillators can be applied to complex optimization and machine learning tasks, such as Ising problems \cite{Utsunomiya2011}.

In most dynamical systems, spontaneous oscillations arise from a Hopf bifurcation \cite{StrogatzBook}.  In optics, the simplest such system is the non-degenerate $\chi^{(2)}$ optical parametric oscillator (OPO), which behaves as a quantum-limited amplifier below threshold \cite{Yamamoto1990} and has a symmetric limit cycle above \cite{Reid1987}.  In addition, cavity quantum electrodynamics (QED) systems can self-oscillate in the right conditions \cite{Armen2006, Kwon2013}.  However, nanofabrication with $\chi^{(2)}$ materials such as KTP and LiNbO$_3$ is still in its infancy \cite{Poberaj2012}, and most implementations of cavity QED -- trapped atoms, quantum dots, NV centers -- are not scalable with current technology.  To realize neuromorphic computing with photonics, there is an unfulfilled need for self-oscillating photonic devices based on a scalable technology.

Free-carrier dispersion can fulfill this unmet need.  This effect is present in silicon and all III-V semiconductors, and is scalable and low-power \cite{Notomi2010}.  Previous work by Malaguti et al.~\cite{Malaguti2011, Malaguti2013} and Chen et al.~\cite{Chen2012} showed that when the photon and carrier lifetime are comparable, an optical cavity can pass through a Hopf bifurcation and undergo self-oscillation.  However, these studies focused on the many-photon classical limit, where quantum fluctuations can be ignored.  If such a device is optimized for low power, quantum fluctuations in the photon and carrier number may substantially alter the dynamics and limit the performance of real devices.

In our previous work \cite{Paper1}, we derived a set of stochastic equations for free-carrier optical cavities that  model these quantum fluctuations.  Here, we apply those equations to study the effects of quantum noise on the free-carrier Hopf bifurcation.

Sections \ref{sec:conditions} and \ref{sec:sims} discuss the general theory of the oscillations, which arise from an instability in the linearized model around the system's fixed point.  Because this is done in a general, scale-invariant way, it should be possible to observe these oscillations in a wide range of systems spanning orders of magnitude in speed, size and energy.  Next, we consider the equations of motion close to the bifurcation point and show that the bifurcation resembles the non-degenerate OPO at threshold with some extra noise.  Section \ref{sec:07-below} models the device below threshold: it functions as a phase-insensitive linear amplifier with noise $\sim 5\times$ above the Caves bound \cite{Caves1982}.  The near-threshold behavior, which follows the critical exponents of the Hopf bifurcation, is discussed in Section \ref{sec:near}.

The above-threshold case is covered in Section \ref{sec:above}.  Like the non-degenerate OPO, the free-carrier cavity has a limit cycle in this regime.  The above-threshold OPO can be considered a ``quantum-optimal'' limit cycle in the sense that it can function as an optimal homodyne detector.  By comparison, the free-carrier limit cycle is $\sim 10\times$ noisier than the OPO.  This difference is due to the incoherent nature of carrier excitation and decay.

Limit-cycle devices can be very useful in optimization and machine learning.  In Section \ref{sec:ising}, we propose and simulate an Ising machine based on the free-carrier limit cycle, which should be several orders of magnitude faster and less power-consuming than a supercomputer.  In addition, Section \ref{sec:relay} discusses an all-optical XOR gate based on the limit-cycle effect.

\section{Conditions for Self-Oscillation}
\label{sec:conditions}

\subsection{Equations of Motion}

A single-mode free-carrier optical cavity has three degrees of freedom: two field quadratures $(\alpha, \alpha^*)$ and the free carrier number $N$.  Typically, the following effects are relevant:

\begin{enumerate}
	\item Cavity-waveguide coupling.  This gives rise to a linear loss $\kappa$ in the cavity field.
	\item Linear and two-photon absorption.  The former dominates for near-bandgap operation of direct-gap semiconductors; the latter for indirect-gap systems.  Gives rise to a linear loss term $\eta$ and a quadratic loss term $\beta$.  Both act as source terms for the carrier number.
	\item Free-carrier dispersion / absorption.  The cavity detuning shifts as a function of the carrier number: $\Delta \rightarrow \Delta + \delta_c N$.  If $\delta_c = \delta_1 - i\,\delta_2$ is complex, this accounts for free-carrier absorption as well.
	\item Carrier decay.  Typically due to recombination at surface sites or diffusion out of the cavity.  This gives rise to a linear loss term $\gamma$ for $N$.
\end{enumerate}

In this text, we ignore the following effects:

\begin{enumerate}
	\item Excitons, which tend to be the dominant effect only at low temperatures or in exotic materials.
	\item Thermo-optic effect.  Temperature changes much more slowly than the photon or carrier number, so does not typically play a role in the fast dynamics of the device.  It may, however, lead to stability issues, which are not the focus of this paper \cite{VanVar2012, JohnsonThesis, Johnson06}.
	\item Optomechanical effects, which are negligible unless a cavity has been specifically engineered to probe them.
\end{enumerate}

Under these assumptions, the device can be modeled as an open quantum system that couples to a Markovian bath; see generally \cite{Gardiner1985, WallsMilburn, QuantumNoise}.  The full quantum theory is quite involved and is discussed in our previous paper \cite{Paper1}.  In short, starting from a quantum model with a bosonic photon mode and many fermionic carrier modes, one can construct a generalized Wigner function in terms of a set of bosonized operators and derive a Fokker-Planck equation for this function using the truncated Wigner method \cite{Santori2014, Gronchi1978}.  This can be recast as a set of stochastic differential equations (SDE's) which sample from the Wigner function as a probability distribution.  Assuming that dephasing and thermalization are much faster than the photon or carrier lifetimes, one obtains the following stochastic equations of motion (Eqs.~(C16-17) from \cite{Paper1}):

\begin{widetext}

\begin{eqnarray}
    d\alpha & = & \left[-\frac{\kappa+\eta}{2} - (\beta+i\chi)\alpha^*\alpha - i(\Delta + N \delta_c)\right]\alpha\,dt - \sqrt{\kappa}\, d\binA + \underbrace{\left[-\sqrt{\eta}\, d\beta_\eta - 2\sqrt{\beta}\alpha^*d\beta_\beta - \sqrt{2N \delta_2}\, d\beta_{fca}\right]}_{d\xi_\alpha} \label{eq:07-eom1} \\
    dN & = & \left[\eta\,\alpha^*\alpha + \beta(\alpha^*\alpha)^2 - \gamma N\right]dt + \underbrace{\left[\sqrt{\eta}(\alpha^*d\beta_\eta + \alpha\,d\beta_\eta^*) + \sqrt{\beta}\bigl((\alpha^*)^2 d\beta_\beta + \alpha^2 (d\beta_\beta)^*\bigr) + \sqrt{\gamma N} dw_\gamma\right]}_{d\xi_N} \label{eq:07-eom2}
\end{eqnarray}
and the output optical field is:

\beq
	d\boutA = -\sqrt{\kappa}\,\alpha\,dt + d\binA
\eeq

In these equations, $d\binA$ is a complex Wiener process representing the input field, which for vacuum input has the Ito rule $d\binA d\binA^* = dt/2$.  The processes $d\beta_\eta$, $d\beta_\beta$ and $d\beta_{fca}$ correspond to linear, two-photon and free-carrier absorption respectively, and also have vacuum statistics.  The $dw_\gamma$ is a real Wiener process satisfying $dw_\gamma^2 = dt$, giving the Poisson statistics of carrier decay.  The real and imaginary parts of $\delta_c$ are $\delta_c = \delta_1 - i\delta_2$.  Typical values for the parameters in (\ref{eq:07-eom1}-\ref{eq:07-eom2}) are given in Table \ref{table:t1}.

These equations resemble the coupled-mode equations used to analyze semiconductor microcavities elsewhere in the literature \cite{Malaguti2011, Malaguti2013, Chen2012}.  Unlike the equations used elsewhere, (\ref{eq:07-eom1}-\ref{eq:07-eom2}) include quantum-noise terms.  As a result, these equations allow us to model the quantum behavior of devices previously only discussed classically, and study the fundamental quantum limits to device performance.

We can analyze optical bistability and self-oscillation by linearizing these equations of motion about their equilibrium point.  Defining the doubled-up vector $\bar{x} = (\delta\alpha,\delta\alpha^*,\delta N)$, the equations of motion take the following form:

\bea
    \underbrace{d\begin{bmatrix} \delta\alpha \\ \delta\alpha^* \\ \delta N \end{bmatrix}}_{d\bar{x}} & = &
    \underbrace{\begin{bmatrix} -\frac{\eta+\kappa}{2} - i(\Delta + N\delta_c) - 2(\beta+i\chi)\alpha^*\alpha & -(\beta+i\chi)\alpha^2 & -i\delta_c\alpha \\ -\bigl((\beta+i\chi)\alpha^2\bigr)^* & \bigl(-\frac{\eta+\kappa}{2} - i(\Delta + N\delta_c) - 2(\beta+i\chi)\alpha^*\alpha\bigr)^* & (-i\delta_c\alpha)^* \\ (\eta + 2\beta\alpha^*\alpha)\alpha^* & (\eta + 2\beta\alpha^*\alpha)\alpha & -\gamma \end{bmatrix}\ \begin{bmatrix} \delta\alpha \\ \delta\alpha^* \\ \delta N \end{bmatrix} dt}_{\bar{A} \bar{x}\,dt} \nonumber \\
    & & + \underbrace{\begin{bmatrix} -\sqrt{\kappa} & 0 \\ 0 & -\sqrt{\kappa} \\ 0 & 0 \end{bmatrix} \begin{bmatrix} d\binA \\ d\binA^* \end{bmatrix}}_{\bar{B}\,d\bbinA}
    + \underbrace{\begin{bmatrix}
    	-\sqrt{\eta}\,d\beta_\eta - 2\sqrt{\beta}\alpha^*d\beta_\beta - \sqrt{2N \delta_2}\, d\beta_{fca} \\
    	-\sqrt{\eta}\,d\beta_\eta^* - 2\sqrt{\beta}\alpha\, d\beta_\beta^* - \sqrt{2N \delta_2}\, d\beta_{fca}^* \\
		\sqrt{\eta}(\alpha^*d\beta_\eta + \alpha\,d\beta_\eta^*) + \sqrt{\beta}((\alpha^*)^2 d\beta_\beta + \alpha^2 (d\beta_\beta)^*) + \sqrt{\gamma N} dw_N \end{bmatrix}}_{\bar{F} dw} \label{eq:07-linabcd}
\eea
\end{widetext}

Likewise, the output can be related to the input and internal state by:

\beq
	\underbrace{\begin{bmatrix} d\boutA \\ d\boutA^* \end{bmatrix}}_{d\bboutA} =
	\underbrace{\begin{bmatrix} \sqrt{\kappa} & 0 & 0 \\ 0 & \sqrt{\kappa} & 0 \end{bmatrix}
		\begin{bmatrix} \delta\alpha \\ \delta\alpha^* \\ \delta N \end{bmatrix} dt}_{\bar{C}\bar{x}dt} +
	\underbrace{\begin{bmatrix} 1 & 0 \\ 0 & 1 \end{bmatrix}
    	\begin{bmatrix} d\binA \\ d\binA^* \end{bmatrix}}_{\bar{D}d\bbinA} \label{eq:07-linabcd2}
\eeq
Together, Eqs.~(\ref{eq:07-linabcd}-\ref{eq:07-linabcd2}) may be written formally as:

\bea
	d\bar{x} & = & \bar{A}\bar{x}\,dt + \bar{B}d\bbinA + \bar{F}dw \label{eq:abcd-1} \\
	d\bboutA & = & \bar{C}\bar{x}\,dt + \bar{D}d\bbinA \label{eq:abcd-2}
\eea
which is the standard form for a linear stochastic input-output system.

Equation (\ref{eq:07-linabcd}) separates the dynamics into three parts: a deterministic term $\bar{A}\bar{x}dt$, noise due to quantum fluctuations of the input $\bar{B}\,d\bbinA$, and additional free-carrier noise $\bar{F}dw$.  (Here, $dw$ is a vector Wiener process constructed from the real and imaginary parts of the noise terms $d\beta_\eta, d\beta_\beta, d\beta_{fca}, dw_\gamma$, and normalized to satisfy the Ito table $dw_i dw_j = \delta_{ij}dt$; the matrix $\bar{F}$ is constructed so that (\ref{eq:07-linabcd}) is satisfied).

The matrix $\Amat$ has three eigenvalues.  Due to its doubled-up structure, complex eigenvalues must come in conjugate pairs.  Thus, $\Amat$ can either have three real eigenvalues or one real eigenvalue and one complex conjugate pair.  If the equilibrium is stable, all three eigenvalues must have a negative real part.

There are two ways for an equilibrium to go unstable.  First, a negative real eigenvalue can cross zero and turn positive.  Since only a single direction goes unstable, the equilibrium point bifurcates into two stable equilibria.  This is the standard cusp catastrophe of optical bistability in Kerr and cavity QED systems \cite{Agrawal1979}.  In our previous paper \cite{Paper1}, we discussed it in the context of carrier-based switches and amplifiers.  By calculating the determinant of $\Amat$, we can catch this instability -- for stable equilibrium, $\det \Amat < 0$, but if the equilibrium transitions to unstable, $\det \Amat$ will become positive.

\textit{Self-oscillation} takes place when a conjugate pair of eigenvalues cross the imaginary axis.  In this case, two directions go unstable, so the equilibrium point bifurcates into a ring of steady states, or more often, a limit cycle.  The determinant will remain negative, but the product

\beq
L(\Amat) \equiv \left(\text{tr}(\Amat)^2 - \text{tr}(\Amat^2)\right)\text{tr}(\Amat) - 2\det(\Amat)
\eeq
changes sign at this bifurcation.  To see why, suppose that the matrix $\Amat$ has eigenvalues $\lambda, \mu, \mu^*$.  Then for some transformation $P$,

\beq
P^{-1} \Amat P  = \begin{bmatrix} \lambda & & \\ & \mu & \\ & & \mu^* \end{bmatrix}
\eeq
By the cyclic property of traces and determinants, $L(\Amat) = L(P^{-1} \Amat P)$, and the latter evaluates to:

\beq
L(\Amat) = L(P^{-1} \Amat P) = 4|\lambda + \mu|^2 \text{Re}(\mu)
\eeq
This will change sign from negative to positive when passing through a Hopf bifurcation.

\begin{table}[b!]
\centering
\begin{tabular}{c|l|cc}
Name & Description & GaAs PhC\footnote{GaAs, $\hbar\omega = 0.9 E_g$, $\tilde{V} = 0.25$, $Q = 5000$, $\tau_{fc} = 2$ ps; see \cite{Paper1}, compare \cite{Nozaki2010}.} & Si $\mu$-ring, \footnote{Si, $\lambda = 1.5\mu$m, $\tilde{V} = 40$, $Q = 4 \times 10^5$, $\tau_{fc} = 3$ ns; see \cite{JohnsonThesis}.} \\ \hline
$k$            & $\kappa+\eta$          & $0.42$ ps$^{-1}$      & $0.31$ ns$^{-1}$ \\
$\kappa$       & I/O Coupling           & $k/2$\footnote{All dimensional quantities in this table are scaled to the linear loss $k$.}                 & $k$ \\
$\eta$         & LA                     & $k/2$                 & $0$ \\
$\beta$        & TPA                    & $7.9 \times 10^{-5}k$ & $3.7 \times 10^{-6}k$ \\
$\chi$         & Kerr                   & $0$\footnote{Negligible, as dispersive effect is dominated by free carriers.}                   & $0$ \\
$\delta$       & FCD                    & $2.7 \times 10^{-3}k$ & $(5.6-0.4i)\times 10^{-4}k$ \\
$\gamma$       & Carrier Decay          & $1.2k$                & $1.0k$ \\ \hline
$\bar{\delta}$ & $\delta_2/\delta_1$    & $0$                   & $0.07$ \\
$\bar{\zeta}$  & $\delta_1/\beta$       & $34$                  & $150$ \\
$\bar{\chi}$   & $\chi/k$               & $0$                   & $0$ \\
$\bar{\gamma}$ & $\gamma/k$             & $1.2$                 & $1.2$ \\
$\bar{\kappa}$ & $\kappa/k$             & $0.5$                 & $1.0$ \\
$\bar{\eta}$   & $\eta/k$               & $0.5$                 & $0$ \\
$\bar{\Delta}$ & $\Delta/k$             & varies                & varies \\ \hline\hline
\end{tabular}
\caption{Free-carrier cavity parameters for a state-of-the art GaAs photonic crystal and Si microring cavity.}
\label{table:t1}
\end{table}

\subsection{Scaling Laws}
\label{sec:07-scaling}

\begin{figure}[tbp]
\begin{center}
\includegraphics[width=1.00\columnwidth]{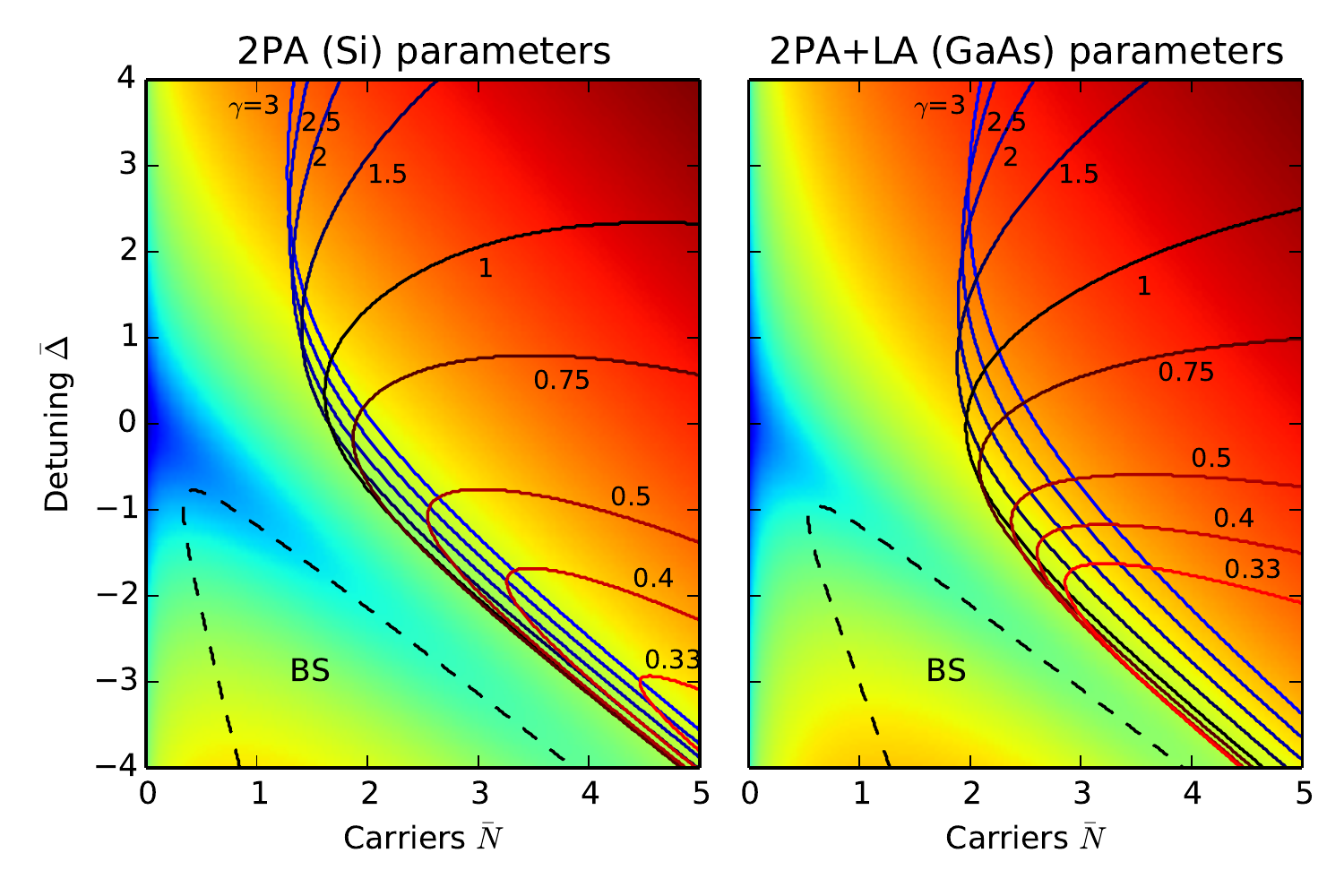}
\caption{Oscillation region as a function of cavity parameters.  Two materials are shown: Si at 1.5 $\mu$m (left) and GaAs near the band edge (right).  Oscillations occur to the right of the solid curves.  Curves represent different values of $\bar{\gamma}$, from 0.33 to 3.0.  Optical bistability occurs in the dashed region.  Color represents the steady-state input power.}
\label{fig:07-f1}
\end{center}
\end{figure}

Equations (\ref{eq:07-eom1}-\ref{eq:07-eom2}), and the resulting matrix $\Amat$, have 8 free parameters.  That's a lot.  Naively, searching for oscillating conditions would appear difficult because of all the parameters one must consider.  However, several scaling laws let us reduce this to 6 ``normalized'' parameters, of which 3 are material constants.

Start with equations of motion (\ref{eq:07-eom1}-\ref{eq:07-eom2}).  Let $k = \kappa + \eta$ be the total cavity linear loss.  Scale time, the electric field, the input field, and the carrier number as follows:

\beq
	t \rightarrow \frac{\bar{t}}{k},\ \ \
	\alpha \rightarrow \frac{\bar{\alpha}}{\sqrt{\beta/k}},\ \ \
	\binA \rightarrow \frac{\bar{\beta}_{\rm in}}{\sqrt{\beta/k^2}},\ \ \
	N \rightarrow \frac{\bar{N}}{\delta_1/k} \nonumber
\eeq

Intuitively, time $\bar{t}$ is scaled so that the cavity photon lifetime is one.  The carrier number is scaled so that $\bar{N} = 1$ shifts the cavity by one linewidth.  The intracavity field $\bar{\alpha}$ and input field $\bar{\beta}_{\rm in}$ are scaled to the two-photon absorption: $|\bar{\alpha}| = 1$ means that the single- and two-photon loss processes are equally strong.

The reduced equations take the following form:

\begin{eqnarray}
d\bar{\alpha} & = & \left[-\left(1/2+\bar{\delta}\bar{N}\right) - \bar{\alpha}^*\bar{\alpha} - i\left(\bar{\Delta}+\bar{N}\right)\right]\bar{\alpha}\,d\bar{t} \nonumber\\
& &- \sqrt{\bar{\kappa}}\bar{\beta}_{\rm in} d\bar{t} + \frac{\sqrt{\beta} F_\alpha}{k} d\bar{w} \\
d\bar{N} & = & \left[\bar{\eta}\bar{\zeta} (\bar{\alpha}^*\bar{\alpha}) + \bar{\zeta} (\bar{\alpha}^*\bar{\alpha})^2 - \bar{\gamma} \bar{N}\right]d\bar{t} + \frac{\delta_1 F_N}{k^{3/2}} d\bar{w}
\end{eqnarray}
In the absence of noise, these equations have 6 independent parameters:

\begin{align}
	\bar{\delta} = \frac{\delta_{2}}{\delta_1},\ \
	\bar{\zeta} = \frac{\delta_1}{\beta},\ \
	\bar{\chi} = \frac{\chi}{\beta} \quad
	& \biggr\} \quad \begin{array}{c} {\rm Material} \\ {\rm Properties} \end{array} \nonumber \\
	\bar{\gamma} = \frac{\gamma}{k},\ \
	\bar{\kappa} = 1-\bar{\eta} = \frac{\kappa}{k} \quad
	& \biggr\} \quad \begin{array}{c} {\rm Cavity} \\ {\rm Design} \end{array} \nonumber \\
	\bar{\Delta} = \frac{\Delta}{k} \quad
	& \biggr\} \quad {\rm Tunable}
\end{align}
where $k = \kappa + \eta$ and $\delta_c = \delta_1 - i\delta_2$.

\begin{figure}[tbp]
\begin{center}
\includegraphics[width=1.00\columnwidth]{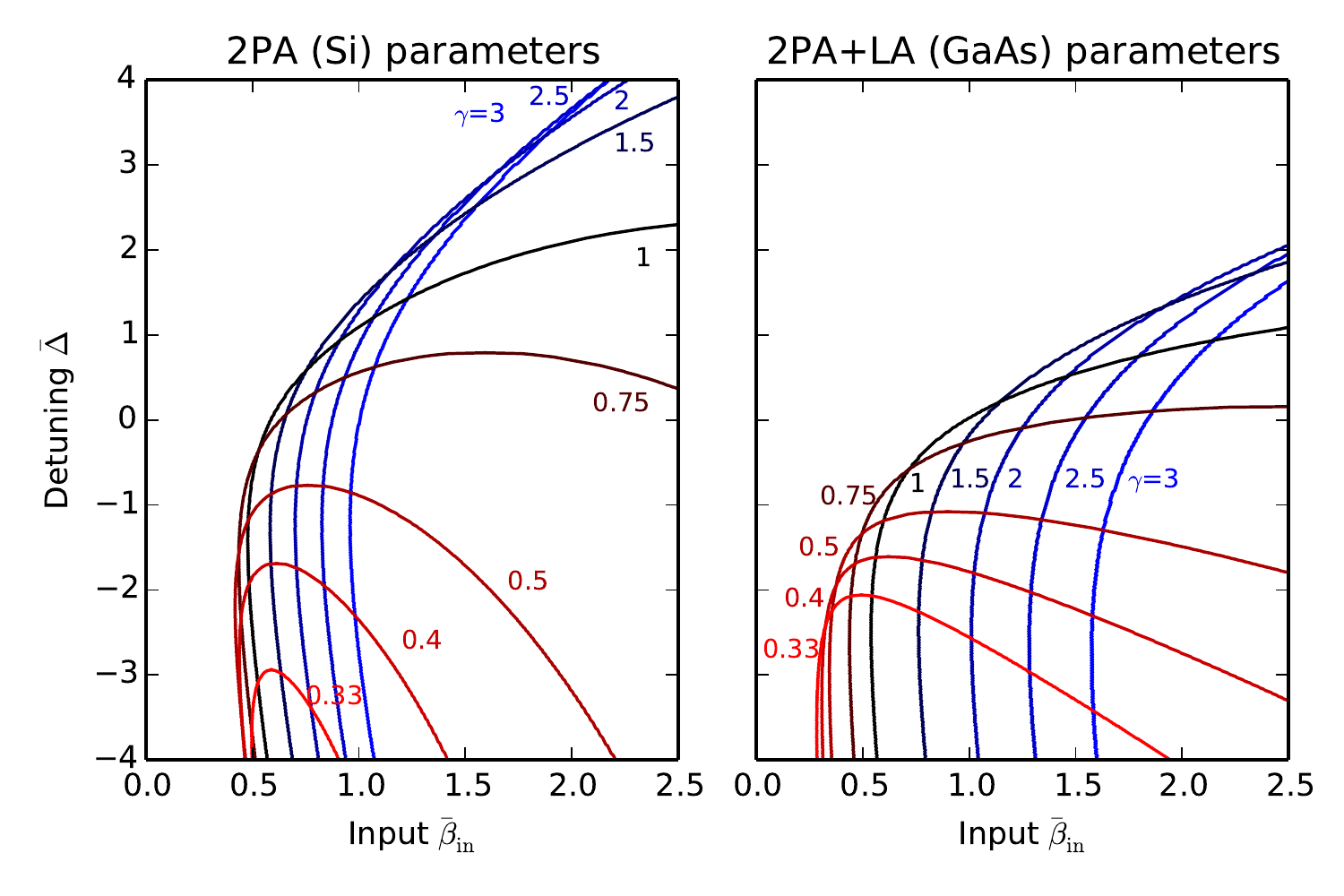}
\caption{Oscillation region as a function of cavity parameters.  Here, the x-axis is normalized input field rather than normalized $N$.}
\label{fig:07-f2}
\end{center}
\end{figure}

Once a material and laser wavelength are picked, only three parameters can be varied.  The relative linear absorption $\bar{\eta} = 1 - \bar{\kappa}$ typically cannot vary much -- in a linear-absorption cavity it should be $O(1)$ to maximize the nonlinearity, and in TPA materials like silicon it is zero.  The ratio of optical to free-carrier lifetimes, $\bar{\gamma}$, can vary by several orders of magnitude, depending on the cavity geometry and $Q$.  For instance, it is easy to make low-$Q$ cavities with a very small $\bar{\gamma}$.  State-of-the-art micro-rings have $Q \sim 10^6$ and $\tau_c \sim $ns and consequently $\gamma/k \sim 1$.  Coincidentally, photonic crystals tend to have a similar ratio, though the carrier decay mechanism (diffusion) is different.  It is also possible to make large cavities with very high $Q$ and large $\bar{\gamma}$.

Obviously, both the input power and detuning can also be varied.  For a given material, these quantities exhaust the parameter space.  By plotting the self-oscillating regions as a function of $\bar{\Delta}$ and $\bar{N}$ (a function of the input), for reasonable values of $\bar{\gamma}$, we are essentially plotting the entire parameter space.  As shown in Figure \ref{fig:07-f1}, in a large fraction of the parameter space, the cavity should self-oscillate.

Figure \ref{fig:07-f2} shows the self-pulsing region as a function of input field and detuning.  This is generally similar to Figure \ref{fig:07-f1}, although the low-$\gamma$ regions appear more accessible because, although the internal carrier number is high, the carriers are long-lived and the cavity requires less optical power.  However, these cavities are complicated by optical bistability (which occurs in the same region), and the slow response time is generally not desirable.  The most desirable conditions seem to occur when the photon and carrier lifetimes are comparable, and the cavity is driven with a slightly detuned pump.

\begin{figure}[tbp]
\begin{center}
\includegraphics[width=1.00\columnwidth]{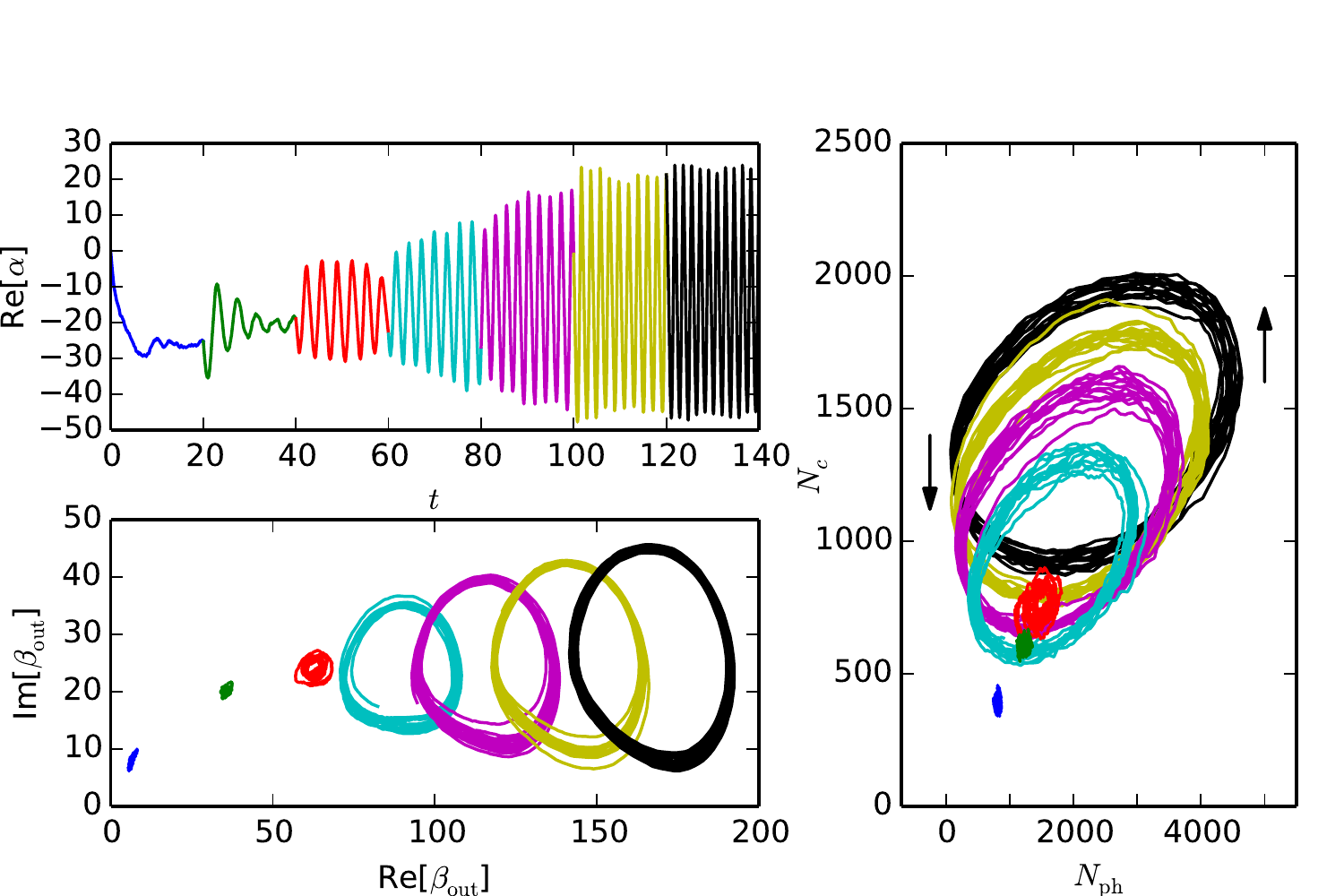}
\caption{Top: time trace of Re[$\alpha(t)$] as the input field is stepped from $\binA = 25,50,75,100,125,150,175$.  Bottom: Output field quadratures at these input powers.  Right: Oscillation between photons and carriers.}
\label{fig:07-f3}
\end{center}
\end{figure}

\section{Semiclassical Simulations}
\label{sec:sims}

Quantum simulations (in the semiclassical Wigner picture) add noise to this model.  For concreteness, in this section and the sections that follow, we consider a GaAs photonic-crystal cavity with parameters given in Table \ref{table:t1}; however, our results are applicable to a range of devices.  Quantities with units of time or inverse time ($t$, $\Delta$, etc.) will be normalized to the cavity lifetime $1/k$.

Figure \ref{fig:07-f3} shows simulations for a detuning $\Delta = -0.8$.  The input field is stepped from $\binA = 25$ (blue) to $175$ (black) in increments of $25$.  The top plot shows a typical time trace.  Oscillations clearly set in at around $\binA = 75$.  In addition to the amplitude, the oscillation frequency also increases with pump power.

The right panel of Figure \ref{fig:07-f3} plots internal photon number (horizontal) against carrier number (vertical).  This provides a qualitative picture of the oscillations: when the photon number is high, more photons are absorbed and the free carrier number increases.  Eventually the carrier number becomes so high that the cavity shifts off-resonance, reducing the cavity's effective driving strength and consequently the photon number.  Once the photon number falls, the carrier number falls because fewer photons are being absorbed, but eventually this brings the cavity back on resonance, increasing the photon number and repeating the cycle.

\begin{figure}[tbp]
\begin{center}
\includegraphics[width=1.00\columnwidth]{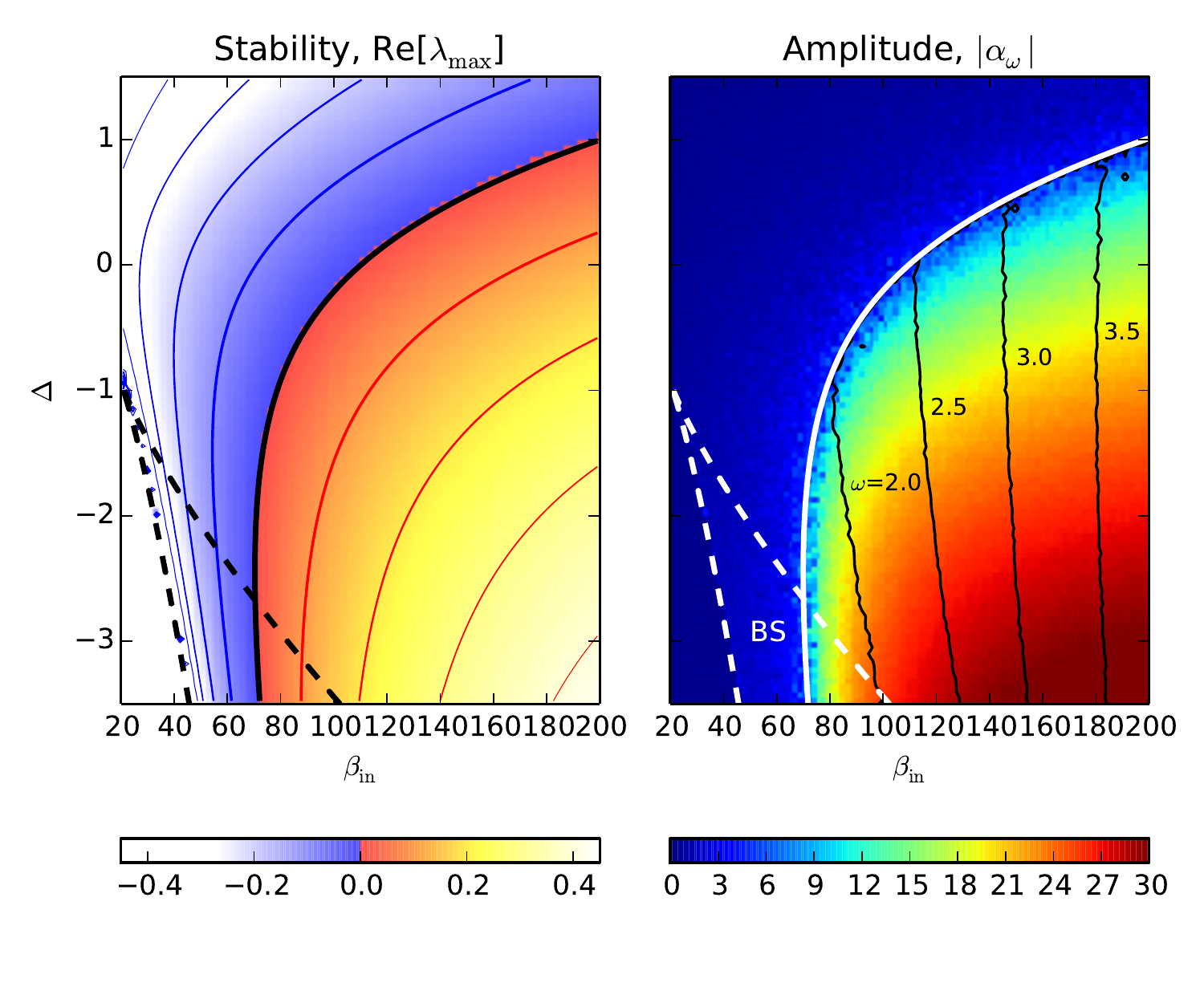}
\caption{Left: Stability of equilibrium point, measured by the real part of the largest eigenvalue of $A$.  Right: Amplitude of limit cycle, with contours designating the limit cycle frequency.}
\label{fig:07-f4}
\end{center}
\end{figure}

To get a more general picture, consider all possible pump powers and detunings for this system.  If a limit cycle forms, we are interested in its amplitude and frequency.  The amplitude should be large, so that a significant fraction of the pump is converted to photons at the limit-cycle frequency.  The frequency should be large enough that the pump and limit cycle fields can be easily demultiplexed with a cavity.  Figure \ref{fig:07-f4} plots both of these figures of merit.  As expected, the amplitude at $\omega$ only becomes nonzero in the unstable region where $\text{Re}[\lambda_{\rm max}] > 0$.  The frequency also grows with pump power, starting at $\omega \approx 1.7$ and growing to $\omega \approx 4$; this is probably a nonlinear effect of the strong pumping.

Two other figures of merit are the limit cycle ``efficiency'' and the gain.  Efficiency is defined in terms of the output and absorbed power:

$$
\eta \equiv \frac{P_{\omega,\rm out}}{P_{\omega,\rm out} + P_{\rm abs}}
$$

\begin{figure}[tbp]
\begin{center}
\includegraphics[width=1.00\columnwidth]{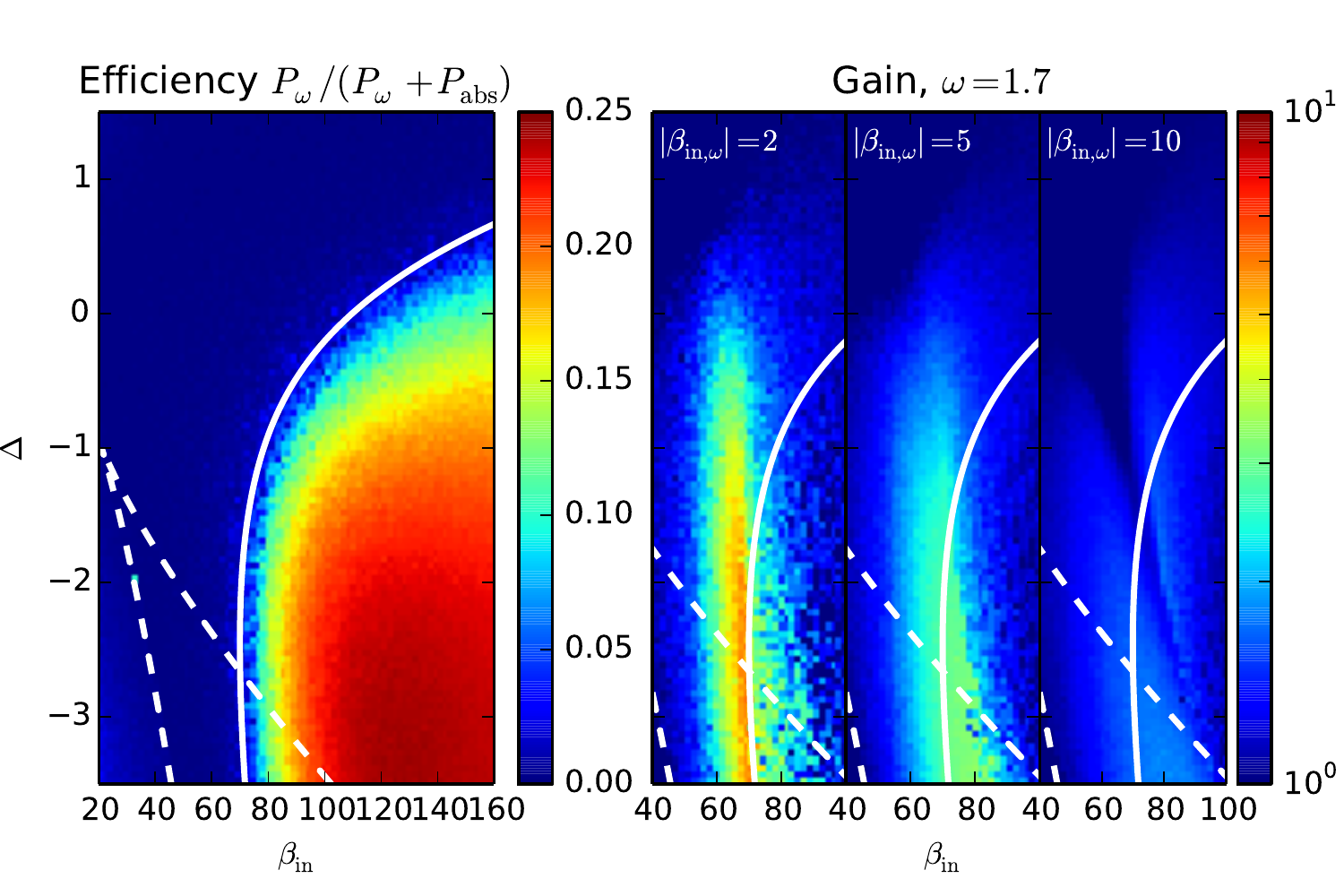}
\caption{Left: Photon conversion efficiency, the ratio of limit cycle photons emitted to photons absorbed.  Right: Amplitude gain $\bout{\omega}/\bin{\omega}$ at $\omega = 1.7$ for different values of seed amplitude $\bin{\omega} = 2, 5, 10$.}
\label{fig:07-f5}
\end{center}
\end{figure}

Efficiency is defined this way rather than output over input because much of the input power is not consumed by the device; it is just a constant bias that can be recycled.  If there finite conversion to $\omega$ and no absorption, we say the efficiency is 1; if no conversion, it is obviously zero.  The left panel of Figure~\ref{fig:07-f5} plots efficiency as a function of detuning and input field.  While not close to 100\%, the efficiency is not too small, either -- peaking at around 20\%.

If we drive the device with a sinusoidal field whose frequency is close to the limit-cycle frequency, that field should be amplified.  In this way, the free-carrier cavity acts as a phase-insensitive amplifier.  The amplitude gain $G(\omega) = \beta_{\omega,\rm out}/\beta_{\omega,\rm in}$ is plotted at $\omega = 1.7$ in the right panel of Figure~\ref{fig:07-f5}.

\section{Below Threshold: Linear Amplification}
\label{sec:07-below}

Below the Hopf bifurcation, a complex pair of eigenvalues approach the imaginary axis.  The corresponding eigenvectors span a plane in phase space; since motion tangent to this plane is only marginally stable, perturbations will be strongly amplified.  Since this plane is two-dimensional, we expect linear, phase-insensitive amplification of both quadratures of the input field \cite{KwonThesis, Wiesenfeld1986}.

For any mesoscopic linear amplifier, an important question to ask is: how much noise does the amplifier have?  Quantum mechanics sets a strict bound on the noise of a quantum linear amplifier \cite{Caves1982}, and this bound is realized with the non-degenerate OPO \cite{Yamamoto1990}.  Since free carriers are excited incoherently, one expects an amplifier driven by carriers to be noisier than a quantum-limited amplifier; however, if the difference is not too large, the free-carrier amplifier may still be preferred because of material, power, or footprint considerations.

\subsection{Nondegenerate OPO}
\label{sec:ndopo}

Although this paper is about free-carrier effects, we introduce the non-degenerate OPO here as a ``benchmark'' system because it is a well-studied system that saturates the Caves bound.  It can be modeled as a quantum input-output system \cite{GoughJames2008, Gardiner1985} with three fields: signal $a$, idler $b$ and pump $c$.  The internal Hamiltonian is

\beq
	H = \Delta_a a^\dagger a + \Delta_b b^\dagger b + \Delta_c c^\dagger c + \frac{\epsilon^*a b c^\dagger - \epsilon a^\dagger b^\dagger c}{2i}
\eeq
and input-output couplings

\bea
	L_1 & = & \sqrt{\kappa_a} a \nonumber \\
	L_2 & = & \sqrt{\kappa_b} b \nonumber \\
	L_3 & = & \sqrt{\kappa_c} c
\eea

Following the Wigner method of \cite{Santori2014}, one can convert the master equation into a PDE for the Wigner function, and truncating higher-order terms, this PDE becomes a Fokker-Planck equation.  This can then be converted into an SDE, and solving the SDE produces trajectories that sample from the Wigner function \cite{WallsMilburn}.  Adiabatically eliminating the pump field and setting $\Delta_a = -\Delta_b \equiv \Delta$, $\kappa_a = \kappa_b \equiv \kappa$ (symmetric doubly-resonant cavity), one obtains the following equations of motion:

\begin{eqnarray}
	d\alpha_1 & = & \left[\left(-i\Delta - \frac{\kappa + \beta\,\alpha_2^*\alpha_2}{2}\right) \alpha_1 + \epsilon\, \alpha_2^*\right]\,dt \nonumber \\
	& & - \sqrt{\kappa}\, d\bin{1} - \sqrt{\beta}\,\alpha_2^* d\bin{3} \\
	d\alpha_2 & = & \left[\left(i\Delta - \frac{\kappa + \beta\,\alpha_1^* \alpha_1}{2}\right) \alpha_2 + \epsilon\, \alpha_1^*\right]\,dt \nonumber \\
	& & - \sqrt{\kappa}\, d\bin{2} - \sqrt{\beta}\,\alpha_1^* d\bin{3} \\
	d\bout{1} & = & \sqrt{\kappa}\,\alpha_1 dt + d\bin{1} \\
	d\bout{2} & = & \sqrt{\kappa}\,\alpha_2 dt + d\bin{2}
\end{eqnarray}
where $\beta = \epsilon^*\epsilon/\kappa$ is the intrinsic coupling strength of the OPO.

Here, $\alpha_1$ and $\alpha_2$ are the signal and idler, which have the same lifetime but opposite detunings.  The pump does not resonate.  These equations are symmetric with respect to $\alpha_1 \leftrightarrow \alpha_2^*$.  Because of the symmetry, the dynamics can be decomposed into a ``symmetric'' mode $\alpha_+ = (\alpha_1 + \alpha_2^*)/2$ and an ``antisymmetric'' mode $\alpha_- = (\alpha_1 - \alpha_2^*)/2$ (and likewise for the $d\beta_\pm$).  In addition, define $dw_1, dw_2$ as quadratures of the pump noise, $d\bin{3} = (dw_1 + i\,dw_2)/2$.  The equations of motion become:

\bea
    d\alpha_\pm & = & \left[(-i\Delta -\kappa/2 \pm \epsilon) \alpha_\pm - \frac{\beta}{2} \bigl(\alpha_\pm^2 - \alpha_\mp^2)\alpha_\pm^*\right]dt \nonumber \\
    & & - \sqrt{\kappa}\,d\bin{\pm} \mp \frac{1}{2}\sqrt{\beta}\left(\alpha_\pm dw_1 - i \alpha_\mp dw_2\right)
\eea

The symmetric mode $\alpha_+$ has gain (a $+\epsilon$ term) while the antisymmetric mode has additional loss.  As a result, at near- or above-threshold pumping, $\alpha_+$ can become very large, but $\alpha_-$ always stays near zero.  In the weakly coupled case ($\beta \ll 1$), we can throw away the terms that couple $\alpha_+$ and $\alpha_-$ in the equation above, and combine the noise terms, giving:

\bea
    d\alpha_+ & = & \left[(-i\Delta -\kappa/2 + \epsilon) \alpha_+ - \frac{\beta}{2} \left|\alpha_+\right|^2 \alpha_+\right]dt \nonumber \\
    & & - \sqrt{\kappa/2} d\beta_+ - \frac{1}{2}\sqrt{\beta}\alpha_+ dw_1 \label{eq:daplus}
\eea

Linearizing about the fixed point $\alpha_1 = \alpha_2 = 0$, and transforming into the frequency domain, we arrive at the input-output relation:

\begin{align}
	& \bout{1}(\omega) = \underbrace{\frac{\left|(\omega - \Delta) + i\kappa/2\right|^2 + (\epsilon/2)^2}{\left(-(\omega - \Delta) + i\kappa/2\right)^2 + (\epsilon/2)^2}}_{e^{i\phi} \cosh\eta} \bin{1}(\omega) \nonumber \\
	& \qquad + \underbrace{\frac{2(\kappa/2)(\epsilon/2)}{\left(-(\omega - \Delta) + i\kappa/2\right)^2 + (\epsilon/2)^2}}_{e^{i\psi} \sinh\eta} \bin{2}^*(-\omega)
\end{align}

For phase-insensitive amplification, the gain $G$ an noise $S$ at frequency $\omega$ may be defined as:

\bea
	G(\omega) & \equiv & \left|\frac{\bout{1}(\omega)}{\bin{1}(\omega)}\right| \label{eq:g} \\
	S(\omega) & \equiv & \sqrt{2P(\omega)},\ \ P(\omega) = \frac{\langle \bout{1}(\omega)^*\bout{1}(\omega')\rangle}{\delta(\omega-\omega')} \label{eq:s}
\eea
In terms of $\eta$, they are:

\beq
	G(\omega) = \cosh\eta,\ \ S(\omega) = \sqrt{2\cosh^2\eta - 1} \label{eq:gw}
\eeq

Note that this $S(\omega)$ is different from the squeezing spectrum of \cite{WallsMilburn, Gough2009Sq}; rather, it is a measure of the electromagnetic energy at frequency $\omega$.  The squeezing spectrum, by contrast, is a power spectrum of a homodyne measurement.

From (\ref{eq:gw}) one sees that the non-degenerate OPO saturates the Caves bound for phase-insensitive amplifiers \cite{Caves1982}:

\beq
S(\omega) \geq \sqrt{2G(\omega)^2 - 1}
\eeq

\begin{figure}[tbp]
\begin{center}
\includegraphics[width=1.00\columnwidth]{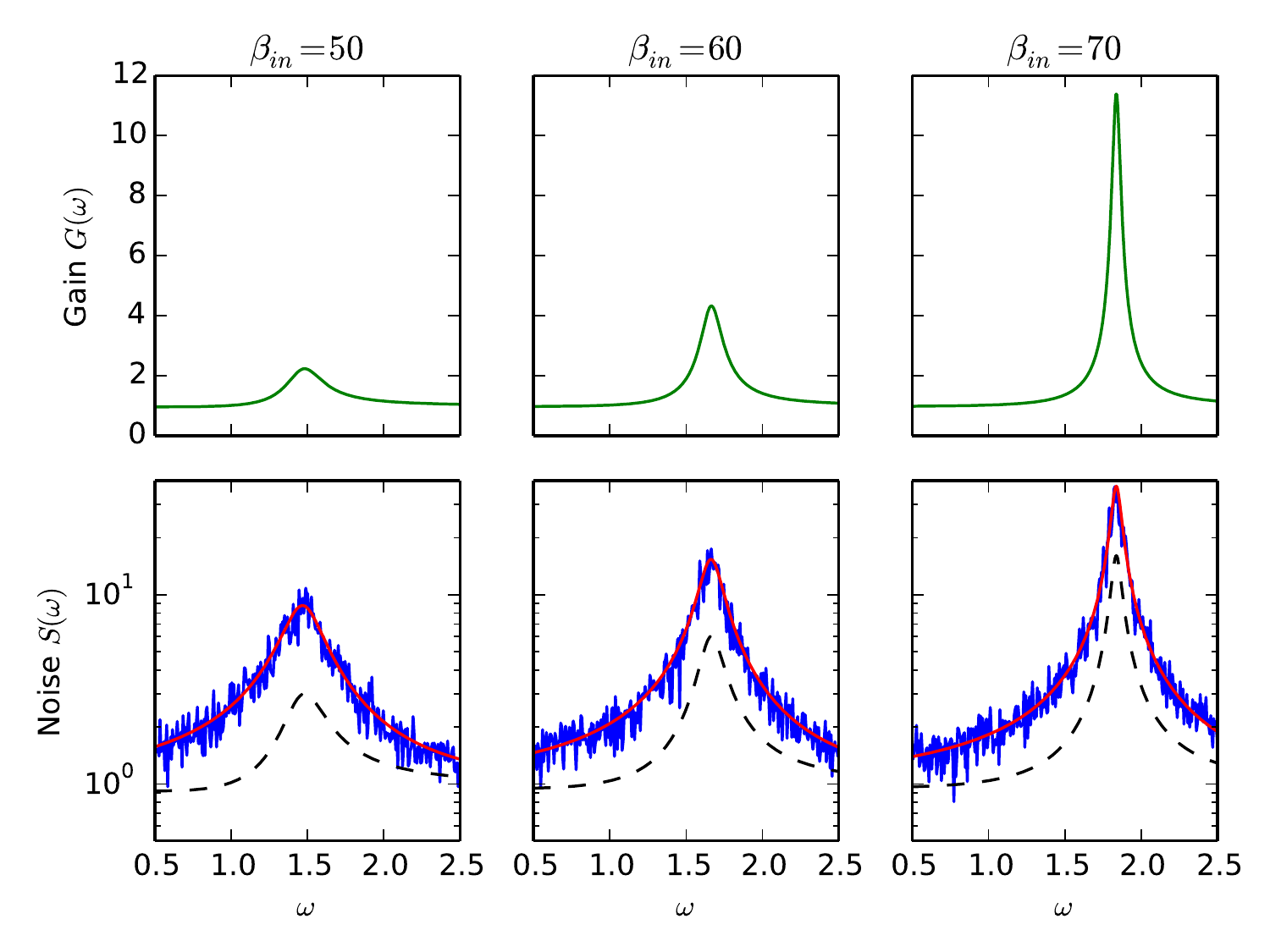}
\caption{Plots of the amplitude gain (top) and noise (bottom) for free-carrier cavity with $\Delta = -1.0$ approaching the Hopf bifurcation.  In the lower graph, the blue line comes from numerical simulation, the red curve is the analytic linearized model, and the black dashed curve is the Caves bound.}
\label{fig:07-f6}
\end{center}
\end{figure}

\subsection{Free-Carrier Amplifier}

Turning to the free-carrier amplifier, first transform Equations~(\ref{eq:abcd-1}-\ref{eq:abcd-2}) to the frequency domain:

\bea
	-i\omega \bar{x}(\omega) & = & \bar{A}\bar{x}(\omega) + \bar{B}\bbinA(\omega) + \bar{F}w(\omega) \\
	\bboutA(\omega) & = & \bar{C}\bar{x}(\omega) + \bar{D}\bbinA(\omega)
\eea
with state $\bar{x}(\omega) = \bigl(\alpha(\omega), \alpha^*(-\omega), N(\omega)\bigr)$ and input-output field $\bar{\beta}(\omega) = \bigl(\beta(\omega), \beta^*(-\omega)\bigr)$.  This is the standard frequency-domain form for doubled-up variables \cite{Gough2010}.

Solving for $\bar{x}$, this becomes a linear input-output relation with a transfer function and a noise matrix:

\bea
	\bboutA(\omega) & = & \underbrace{\left[\bar{D} + \bar{C} \frac{1}{-i\omega - \bar{A}} \bar{B}\right] \bbinA(\omega)}_{T(\omega) \bbinA(\omega)} \nonumber \\
	& & \qquad\ \ + \underbrace{\left[\bar{C} \frac{1}{-i\omega - \bar{A}} \bar{F}\right] w(\omega)}_{N(\omega) w(\omega)} \nonumber
\eea

Applying the definitions of $G$ and $S$ in Eqs.~(\ref{eq:g}-\ref{eq:s}), we find:

\beq
	G(\omega) = |T(\omega)_{11}|,\ \
	\frac{S(\omega)^2}{2} = \left[\frac{T(\omega)T(\omega)^\dagger}{2} + N(\omega)N(\omega)^\dagger\right]_{11}
\eeq

Unlike the OPO, the free-carrier amplifier does not have a simple expression for $G(\omega)$ or $S(\omega)$.  However, they are straightforward to evaluate numerically, and can be compared to a full nonlinear simulation.

Figure \ref{fig:07-f6} shows the gain and noise for the cavity studied in Section \ref{sec:sims}, with $\Delta = -1.0$.  Far from the limit-cycle frequency, there is no gain and the output noise matches that of the vacuum.  As the power is increased and the system approaches the Hopf bifurcation, the gain and noise at the resonance obviously diverge.  But the noise always remains a factor of $~\sim 2-3$ above the Caves bound (in terms of noise power, a factor of $\sim 5$ above the bound).  This is due to the incoherent nature of the free-carrier nonlinearity.

\section{Near Threshold: Critical Exponents}
\label{sec:near}

Near the bifurcation point, the system transitions from a stable fixed point to a stable limit cycle.  Dynamical systems exhibit universal behavior near this bifurcation, in the sense that every system with a Hopf bifurcation can be transformed into the same normal form \cite{StrogatzBook, WigginsBook}.  The same is not true when one adds noise and quantum effects.  Two systems with the same semiclassical equations of motion can behave very differently once quantum noise is added.  Nevertheless, all systems will show the same {\it qualitative} behavior near a bifurcation point.

Before discussing the free-carrier oscillations, consider the non-degenerate OPO near threshold.  Below threshold, there is a stable fixed point at $\alpha_+ = \alpha_- = 0$.  Above threshold, there is a limit cycle at:

\beq
	\left|\alpha_+\right| = \sqrt{\frac{2\epsilon-\kappa}{\beta}}
\eeq

Thus, if we smoothly vary the parameter $\epsilon$ near the bifurcation point, $\epsilon = \kappa/2 + \delta\epsilon$, the limit cycle amplitude goes as $\sqrt{\epsilon}$.  This is a universal feature.  However, not all OPO's are equal up to a transformation -- the behavior of the quantum states depends strongly on the value of $\beta$.  For $\beta \ll 1$, dissipation is dominant and the system stays in a classical state with a positive Wigner function.  For $\beta \gg 1$, the Wigner formalism breaks down.  (This is true for OPO's in general.  It is known that in this regime the {\it degenerate} OPO can access ``highly quantum'' states with non-positive Wigner function such as number states and cat states \cite{Wolinsky1988, Mirrahimi2013, Mabuchi2012}.)

\begin{figure}[tbp]
\begin{center}
\includegraphics[width=1.0\columnwidth]{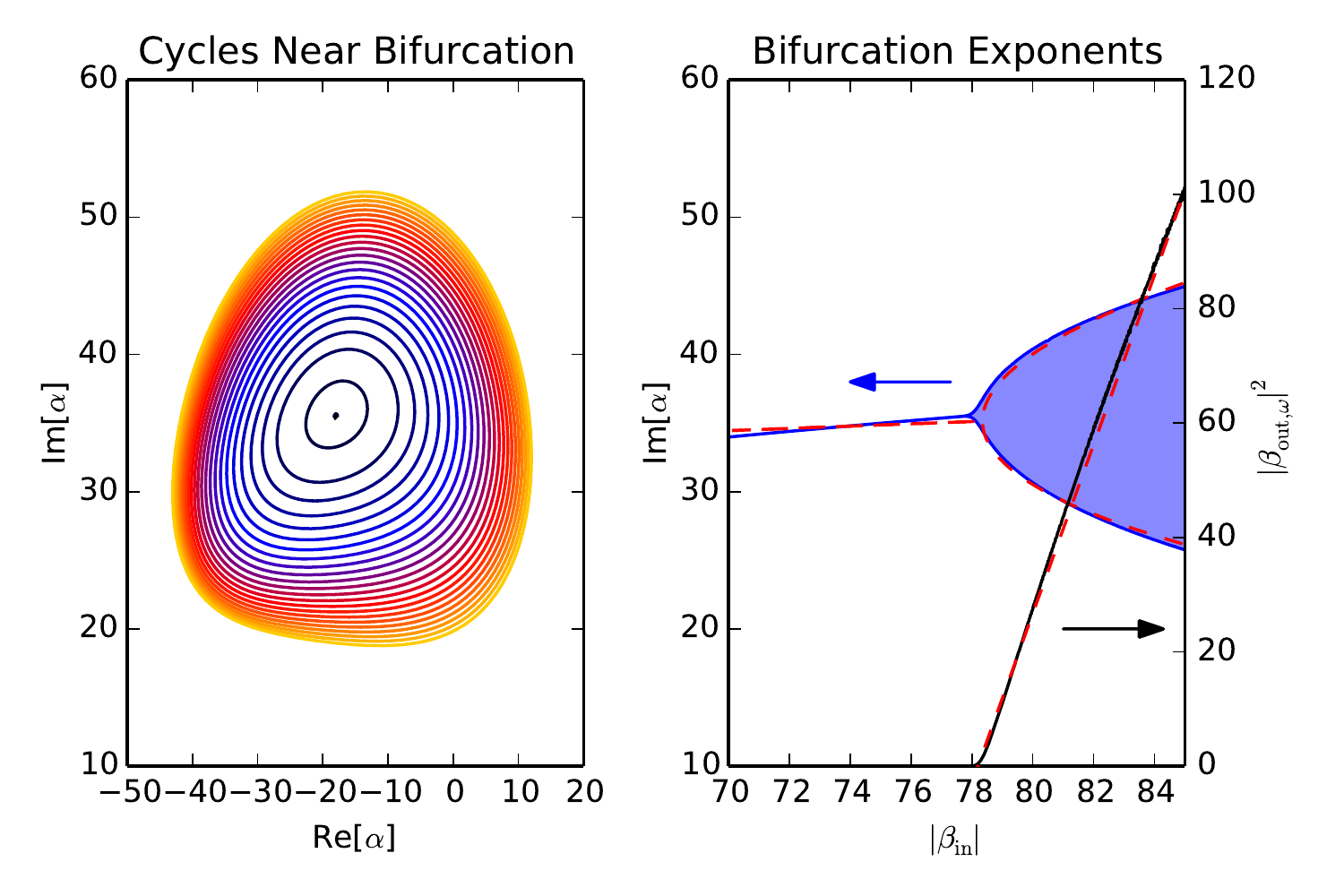}
\caption{Left: Free-carrier limit cycles just above the bifurcation point (noiseless simulation), for evenly spaced $\binA = 78, 79, 80, \ldots$  Right: Size of the limit cycle in terms of $\alpha$ (blue) and $|\boutA(\omega)|^2$ (black), and the critical exponents $\alpha \sim \boutA(\omega) \sim \sqrt{\delta \binA}$}
\label{fig:07-f7}
\end{center}
\end{figure}

The fixed-point eigenvalues near the bifurcation are: $\lambda = (\epsilon - \kappa/2) \pm i\Delta$, and therefore:

\beq
    \left|\alpha_+\right|	\sim \sqrt{\frac{{\rm Re}[\lambda]}{\beta/2}}\ \ \Leftrightarrow \ \
    	\beta \sim \frac{{\rm Re}[\lambda]}{|\alpha_+|^2} \label{eq:07-beta}
\eeq

In classical dynamical systems theory, we can freely transform the system variable $\alpha$, so the parameter $\beta$ can be rescaled to 1.  This is part of the process of transforming to the normal coordinate frame.  Classically, $\alpha$ is dimensional and therefore $\beta$ is not universal in any way.  But in quantum mechanics, there {\it is} a universal scale for $\alpha$: the single-photon scale.  Because of this, $\beta$ becomes a universal parameter, and is related to the ``quantumness'' of the bifurcation.

Figure \ref{fig:07-f7} shows that the free-carrier Hopf bifurcation satisfies the same critical exponent as the non-degenerate OPO: in terms of the input power $\binA$, the average oscillating field goes as $|\alpha| \sim \delta \binA^{1/2}$.  One can calculate the effective $\beta$ for this bifurcation using Eq.~(\ref{eq:07-beta}): fitting to the figures, it works out to $\beta \sim 0.0002$, well in the semiclassical regime.

\begin{figure}[tbp]
\begin{center}
\includegraphics[width=1.0\columnwidth]{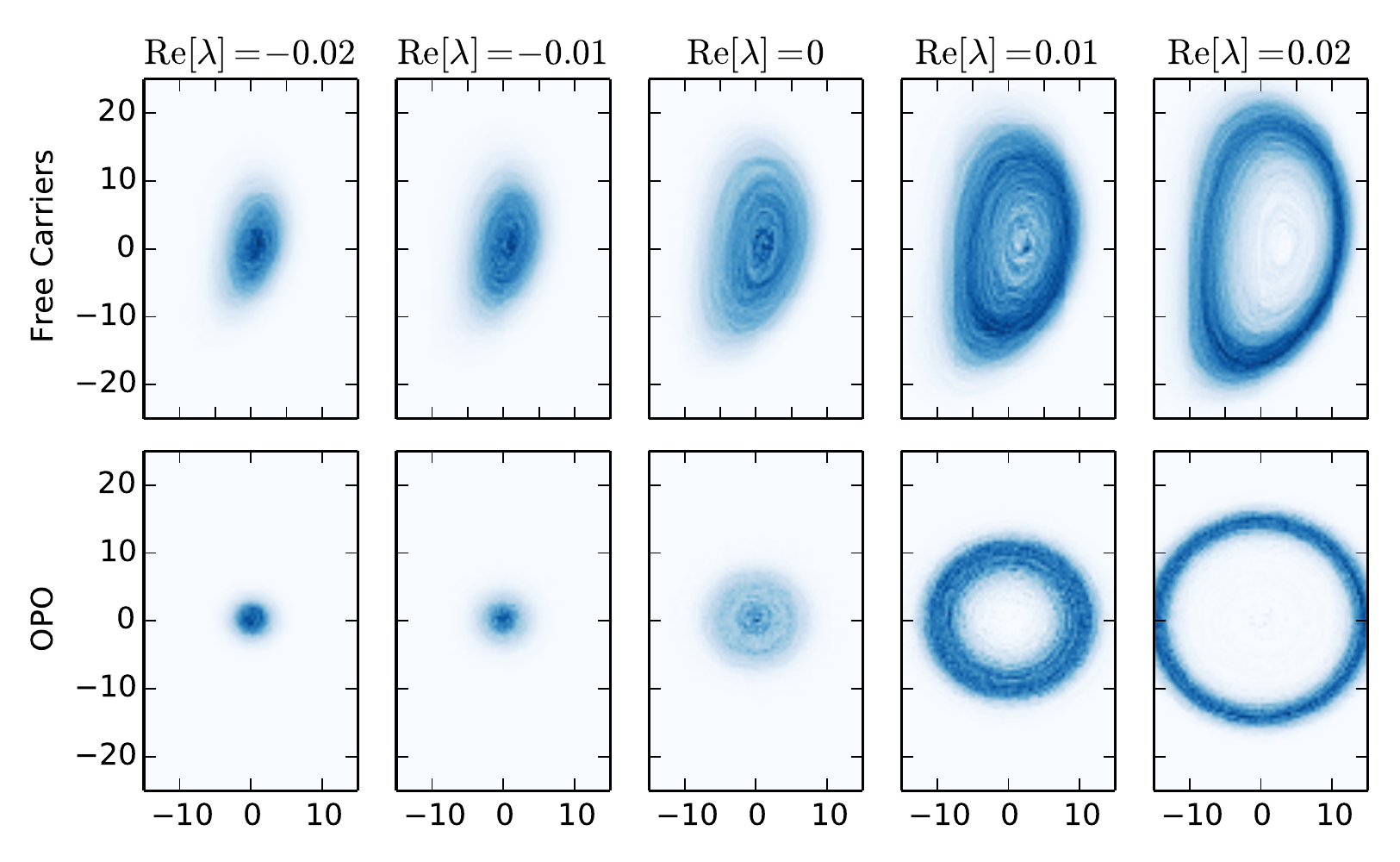}
\caption{Phase plots (axes are Re[$\alpha$], Im[$\alpha$]) of the limit cycles for free carriers ($\Delta = -1.0, a_{\rm in} = 72.5$ through $84.5$) and the non-degenerate OPO ($\beta = 0.0002, \epsilon = 0.48$ through $0.52$).}
\label{fig:07-f8}
\end{center}
\end{figure}

Even after accounting for $\beta$, the free-carrier and OPO Hopf bifurcations are not equivalent up to a transformation, as they would be in classical bifurcation theory.  Again, the culprit is quantum mechanics: the incoherent process of carrier excitation and decay adds extra quantum noise, making the free-carrier limit cycle ``fuzzier'' than its OPO counterpart.  This is shown in Figure~\ref{fig:07-f8}.

\section{Above Threshold: Limit Cycle}
\label{sec:above}

Above threshold, we classically expect a limit cycle.  Quantum noise will blur this out to some degree, but sufficiently far above threshold, the cycle should be clear.

Limit cycles are a classic topic in dynamical systems; some key results are reviewed in Appendix~\ref{sec:04b-lc}.  To summarize the important points:  For an $n$-dimensional phase space, there is a function $(\xi, \vec{u}) \rightarrow \mathbb{R}^n$, that maps the limit cycle phase $\xi$ and local perturbations $\vec{u}$ onto a portion of the phase space.  When the perturbations are small compared to the limit cycle, they can be ignored entirely, reducing the dimensionality of the system from $n$ to 1.  This reduced system has the following equation of motion:

\beq
	d\xi = \omega\,dt + \sum_i{\rm Re}[B_i(\xi)^* d\bin{i}]dt + F(\xi) dw \label{eq:07-lcxi}
\eeq

Here, $B_i(\xi)$ is the response to an external perturbation $d\bin{i}$ and $F(\xi) dw$ is the intrinsic limit cycle noise.

Any limit-cycle system can be used as a homodyne detector.  To see why, consider a coherent input $\bin{i} = \avg{\beta_i}e^{-i\omega_c t} + \bin{i}^{\rm (vac)}$, where $\omega_c$ is the limit cycle frequency.  Averaging over many cycles, this input changes the limit-cycle phase as follows:

\bea
	\Delta\xi - \omega t & = & \int_0^T{\sum_i{\rm Re}[B_i(\xi)^* d\bin{i}]} + \int_0^T{F(\xi) dw} \nonumber \\
	& \sim & N\left(T\sum_i {\rm Re}\left[\mu_{\xi,i} \avg{\beta_i}\right],\ \
		D_\xi T\right) \label{eq:homodyne}
\eea

That is, the phase change has a normal distribution, with mean and variance given by the drift and diffusion constants:

\bea
	\mu_{\xi,i} & = & \avg{B_i(\xi)^* e^{-i\xi}}_\xi \label{eq:mu} \\
	D_\xi & = & \frac{1}{2}\avg{|B_i(\xi)|^2}_\xi + \avg{|F(\xi)|^2}_\xi \label{eq:dxi} \\
	& & \left(\mbox{where}\ \avg{\ldots}_\xi \equiv \frac{1}{2\pi} \int_0^{2\pi} {(\ldots)d\xi}\right) \nonumber
\eea

The drift term $\mu_{\xi,i}$ governs the {\it response rate} of the limit cycle to an external stimulus (in this case, the field).  The diffusion term $D_\xi$ tells us how quickly the limit-cycle phase diffuses in the absence of a stimulus (assuming coherent inputs).  Both terms show up in the homodyne measurement (\ref{eq:homodyne}).  The standard quantum limit \cite{Yamamoto1990} bounds the accuracy of this measurement: in terms of the $\mu_{\xi,i}$ and $D_\xi$, this gives rise to a {\it drift-diffusion inequality}:

\beq
	D_\xi \geq \frac{1}{4} \sum_i |\mu_{\xi,i}|^2 \label{eq:diffeq}
\eeq

This relation holds for all limit cycles.  One can also derive it from Eqs.~(\ref{eq:mu}-\ref{eq:dxi}) by applying the Schwarz inequality.  Equality holds only for special, ``quantum-limited'' limit cycles where $F(\xi) = 0$ and $B_i(\xi) \sim e^{-i\xi}$.  In the sections below, we compare the performance of the non-degenerate OPO and the free-carrier limit cycle using this metric, and show that the OPO saturates the drift-diffusion inequality, while the free-carrier device does not.

\subsection{Non-degenerate OPO}

Again, it will be important to contrast the results obtained here with the non-degenerate OPO; as we will show, this device can function as a quantum-limited homodyne detector for signal and idler fields.  Because it is quantum-limited, no other limit-cycle device will beat the OPO at this task, just like no other linear amplifier can beat the non-degenerate OPO below threshold.

As we show in Appendix~\ref{sec:04b-lc}, the non-degenerate OPO has a limit cycle with $|\alpha_+| = \sqrt{(2\epsilon-\kappa)/\beta}$ and a phase that evolves as:

\bea
	d\xi & = & \Delta\,dt + {\rm Re}\,\left[\frac{-i\sqrt{\kappa}}{\alpha_+}d\bin{+}\right] \nonumber \\
	& = & \Delta\,dt + {\rm Re}\,\left[\frac{-i\sqrt{\kappa}}{2\alpha_+}d\bin{1} + \frac{i\sqrt{\kappa}}{2\alpha_+^*}d\bin{2}\right]
\eea
so that for signal and idler fields varying as $\beta_1 e^{-i\Delta t}$, $\beta_2 e^{i\Delta t}$, the drift-diffusion terms are:

\bea
	\mu_{\xi,1} & = & -i\frac{\sqrt{\kappa}}{2|\alpha_+|} \label{eq:muopo1} \\
	\mu_{\xi,2} & = & -i\frac{\sqrt{\kappa}}{2|\alpha_+|} \label{eq:muopo2} \\
	D_\xi & = & \frac{\kappa}{8|\alpha_+|^2} \label{eq:dopo}
\eea

It is not difficult to see from (\ref{eq:muopo1}-\ref{eq:dopo}) that the drift-diffusion inequality (\ref{eq:diffeq}) is saturated.  In this limit, the non-degenerate OPO functions as an optimal, quantum-limited homodyne detector.

\begin{figure}[tbp]
\begin{center}
\includegraphics[width=0.90\columnwidth]{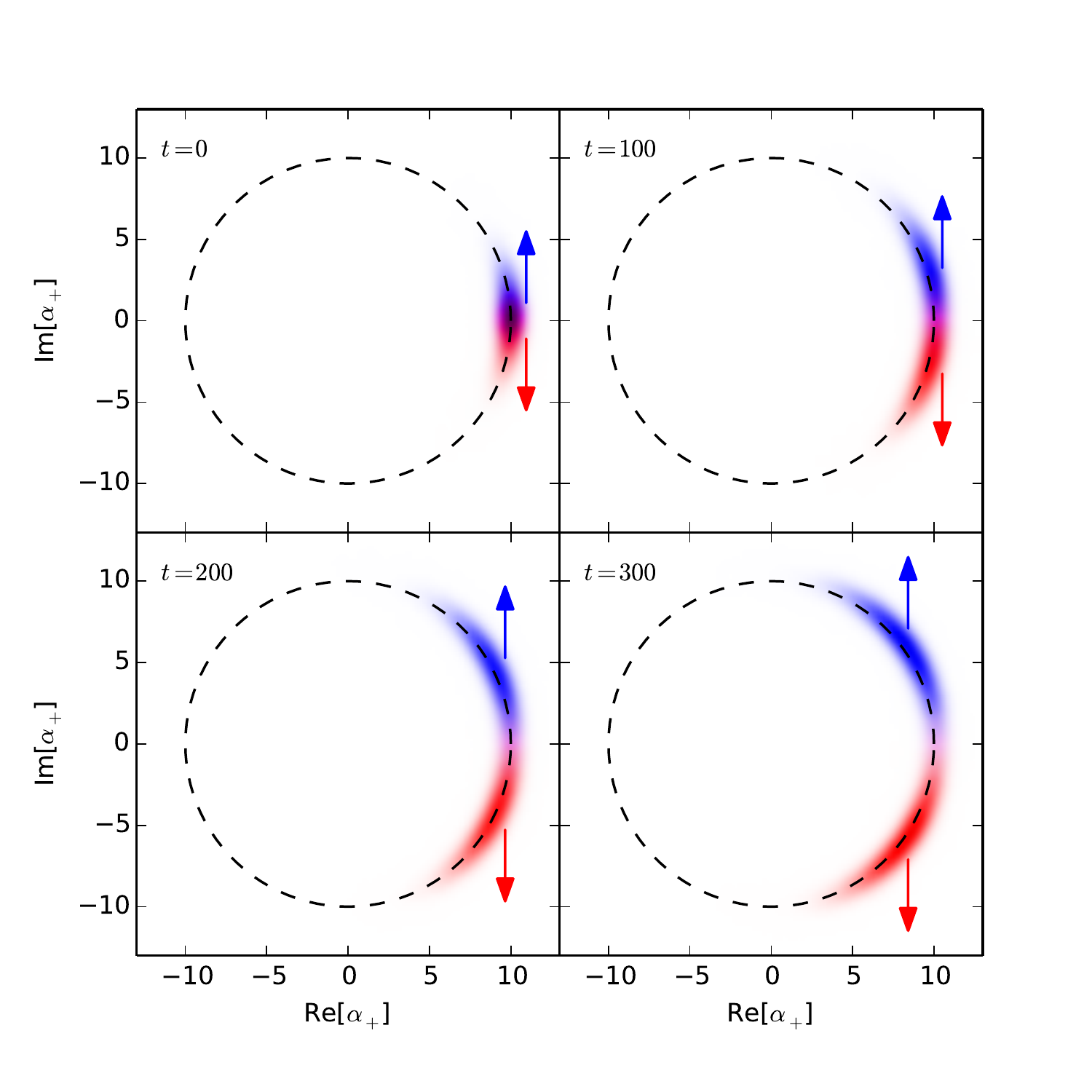}
\caption{Wigner function of the nondegenerate OPO ($\eta = 1.0, \beta = 0.01$) subject to a bias $\beta_1 = 0.15i$ (red) and $-0.15i$ (blue).  The state $\xi(t)$ for $t > 0$, which can be accurately read out with either homodyne or heterodyne detection, effectively encodes a measurement of the $p$-quadrature of the input, ${\rm Im}[\bar{\beta}_1]$.}
\label{fig:04b-f9}
\end{center}
\end{figure}

This is sketched in Figure \ref{fig:04b-f9}.  Here, a non-degenerate OPO with $\Delta = 0$ is used to measure the $p$ quadrature of a signal field.  Depending on the sign of the field, the state either drifts to the top or the bottom, and the diffusion incurred is due to the quantum uncertainty of the homodyne measurement.

\subsection{Free-Carrier Cavity}

Since the equations of motion for the free-carrier cavity are more complicated, a simple analytic expression for $\mu_\xi$ and $D_\xi$ does not exist.  However, these can be computed numerically.  Following the results of Section~\ref{sec:07-below}, it is reasonable to expect diffusion rates 5--10 times faster than for the non-degenerate OPO, the extra diffusion due to incoherent processes involving free carriers.

\begin{figure}[tbp]
\begin{center}
\includegraphics[width=1.0\columnwidth]{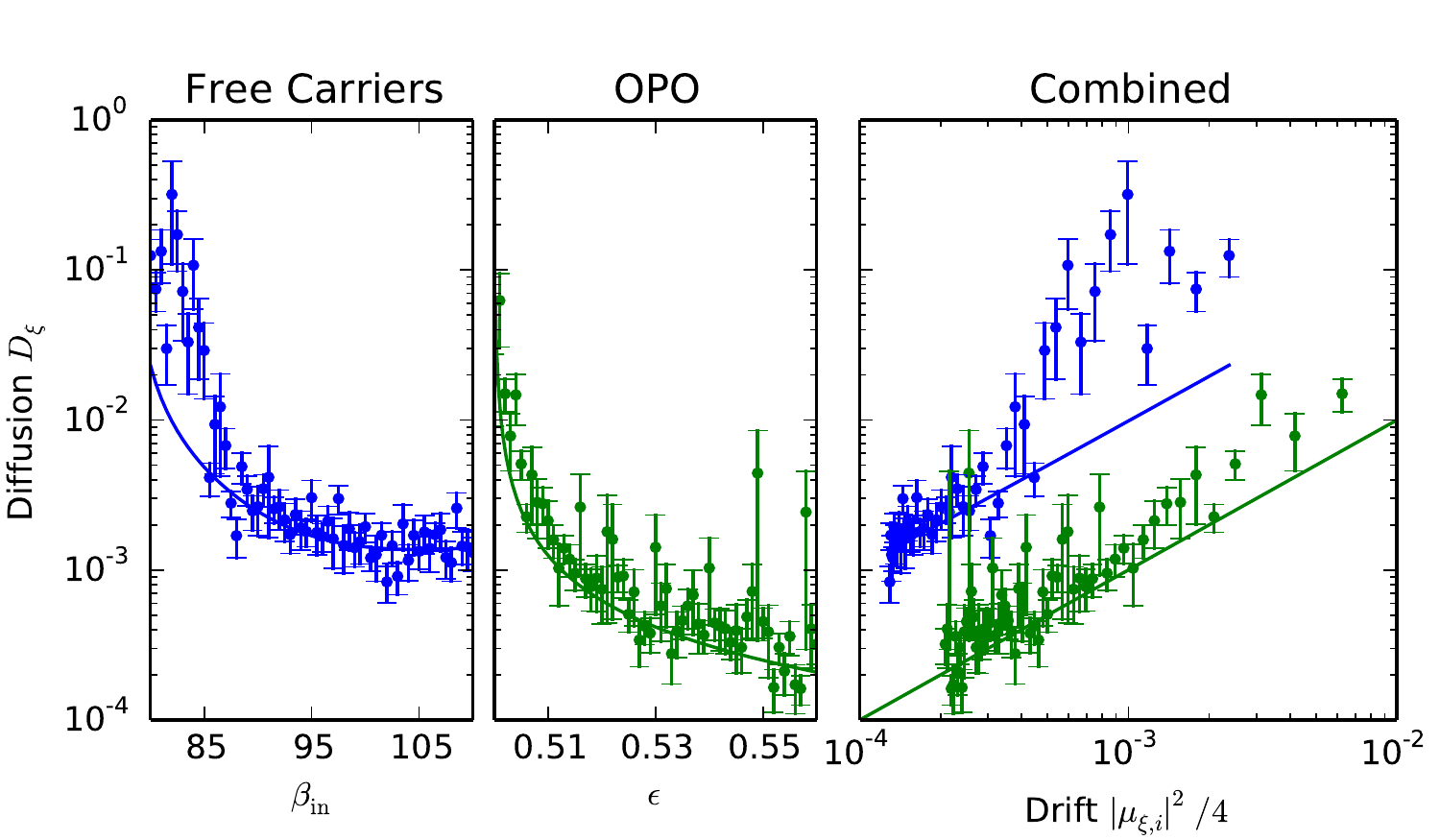}
\caption{Left: Limit-cycle phase diffusion for free-carrier cavity, $\Delta = -1.0$, as a function of input field.  Center: Phase diffusion for non-degenerate OPO, $\beta = 0.0002$, as a function of pump.  Right: Combined, where the drift term $\frac{1}{4}\sum_i |\mu_{\xi,i}|^2$is the common $x$ axis.}
\label{fig:07-f9}
\end{center}
\end{figure}

Figure \ref{fig:07-f9} plots the simulated phase diffusion constant $D_\xi$ for both the OPO and the free-carrier limit cycle.  As one approaches the bifurcation, the diffusion rate increases and diverges from the linearized result (\ref{eq:dxi}), solid curves in the figure.  However, far from the bifurcation, the linearized model agrees with the full simulation for both the OPO and free carriers.

To compare the OPO and free-carrier cavity on equal footing, the right panel of Figure \ref{fig:07-f9} plots the diffusion $D_\xi$ against the right-hand side of (\ref{eq:diffeq}): $\frac{1}{4}\sum_i|\mu_{\xi,i}|^2$.  The OPO simulations, at least for large $|\alpha_+|$, lie on the line $D_\xi = \frac{1}{4}\sum_i|\mu_{\xi,i}|^2$ (green line), while the free-carrier simulations lie a factor of $\sim 10$ above.

\subsection{Entrainment}
\label{sec:driving}

\newcommand{\omegaC}{\omega_c}
\newcommand{\omegaIn}{\omega_{\rm in}}

If the system is driven with a periodic seed field whose frequency $\omegaIn$ does not exactly match the limit-cycle frequency $\omegaC$, the limit cycle may or may not lock to the seed ({\it entrainment}), depending on its amplitude.  To study this effect conceptually, assume a symmetric, noiseless limit-cycle model with a periodic drive $\binA + \bin{\omega} e^{-i\omega t}$, and transform to comoving coordinates $\zeta = \xi - \omegaIn t$.  Equation (\ref{eq:07-lcxi}) takes the form \cite{StrogatzBook}:

\begin{figure}[tbp]
\begin{center}
\includegraphics[width=1.00\columnwidth]{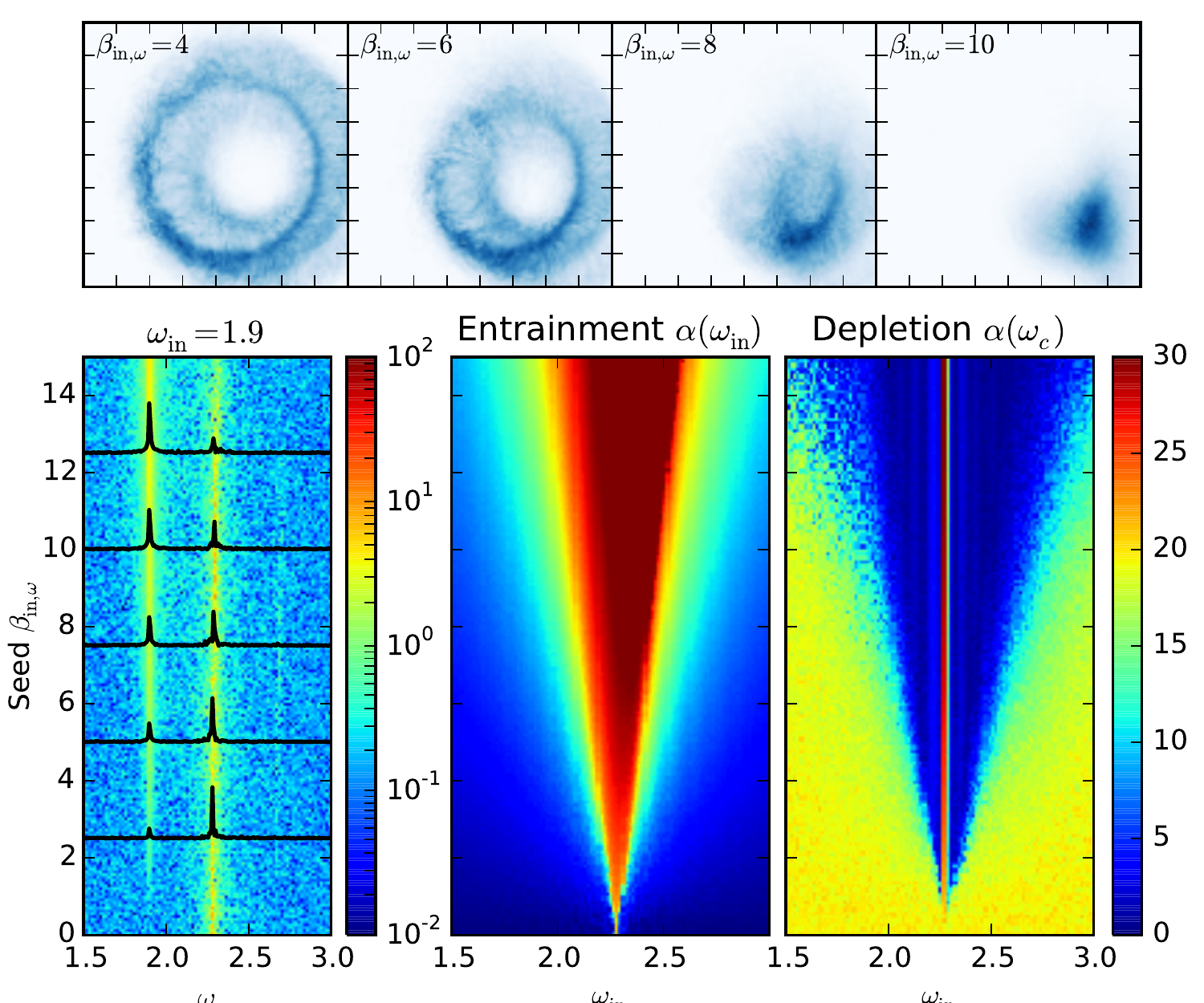}
\caption{Entrainment of free-carrier limit cycle, $\Delta = -1.0$, $\binA = 100$.  Top: Phase plots of the output field in a rotating wave frame, $e^{-i\omegaIn} \boutA$ (mean subtracted).  For large seed inputs, the device clusters to one side of the diagram, indicating phase locking.  Bottom left: output spectrum as a function of seed power, at $\omegaIn = 1.9$.  Bottom right: Entrainment cone.  Plots of $\alpha(\omegaIn)$ and $\alpha(\omegaC)$ (intracavity amplitude at seed and natural frequency, respectively) as a function of seed frequency and amplitude.}
\label{fig:07-f10}
\end{center}
\end{figure}

\beq
	\frac{d\zeta}{dt} = (\omegaC-\omegaIn) - \left|\bin{\omega} B\right| sin(\zeta) \label{eq:07-lczeta}
\eeq

For frequencies $|\omegaC - \omegaIn| < |B \bin{\omega}|$, there is a fixed point at $\zeta = \sin^{-1}((\omegaC-\omegaIn)/|\bin{\omega}B|)$, so the oscillator will lock to the seed.  If we plot $\omegaIn$ on the $x$-axis and $\bin{\omega}$ on the $y$ axis, this phase locking will happen in a vertical cone centered at $(\omegaC, 0)$.  Full free-carrier cavity simulations also show this effect.  Figure~\ref{fig:07-f10} shows results for a $\Delta = -1.0$ cavity with pump $\binA = 100$, which naturally oscillates at $\omegaC = 2.27$.  On top of this, an oscillating field $\bin{\omega} e^{-i\omegaIn t}$ drives the cavity.

The top pane in Figure~\ref{fig:07-f10} shows the real and imaginary quadratures of the output field in a rotating-wave frame: $\tilde{\beta} e^{i\omegaIn t}$.  This is for seed frequency $\omegaIn = 1.9$ and cavity frequency $\omegaC = 2.3$, so $|\omegaIn - \omegaC| \approx 0.4$, or about 16\%.  For weak seed fields, the rotated output makes loops about the origin -- the phase is not locked.  However, around $\bin{\omega} = 10$, it clusters in a given direction -- indicating locking.

The bottom-left plot shows the output spectrum $\boutA(\omega)$ as a function of $\omega$ and the seed amplitude.  One sees two peaks, one at the limit-cycle frequency $\omegaC$ and one at the seed frequency $\omegaIn$.  The peak at the natural frequency $\omegaC$ is strongest when the pump is weak, and eventually goes away for strong pumping.  Conversely, the peak at the drive frequency $\omegaIn$ is absent for weak pumping, and grows with the pump strength.

This is seen more clearly in the bottom-right plots.  Instead of confining ourselves to $\omegaIn = 1.9$, in these plots we vary both the amplitude $\bin{\omega}$ and frequency $\omegaIn$ of the pump.  The left plot shows the power at the input frequency, while the right plot shows the power at the original frequency.  Inside the {\it entrainment cone}, the oscillator locks and the former dominates; outside the cone, the oscillator is unable to lock and the natural frequency is dominant.

From the shape of the entrainment cone, we estimate $B \approx 0.04$ for this set of parameters.

\subsection{Impulse Response}

Suppose that the oscillator has been locked to an external field and now the phase of that field is changed.  The oscillator should follow that phase, but there will be a time lag.  From Eq.~(\ref{eq:07-lczeta}) we can estimate this time lag to be of order:

\beq
	\tau \sim \frac{1}{|\bin{\omega} B|} \label{eq:07-tau}
\eeq

\begin{figure}[tbp]
\begin{center}
\includegraphics[width=1.00\columnwidth]{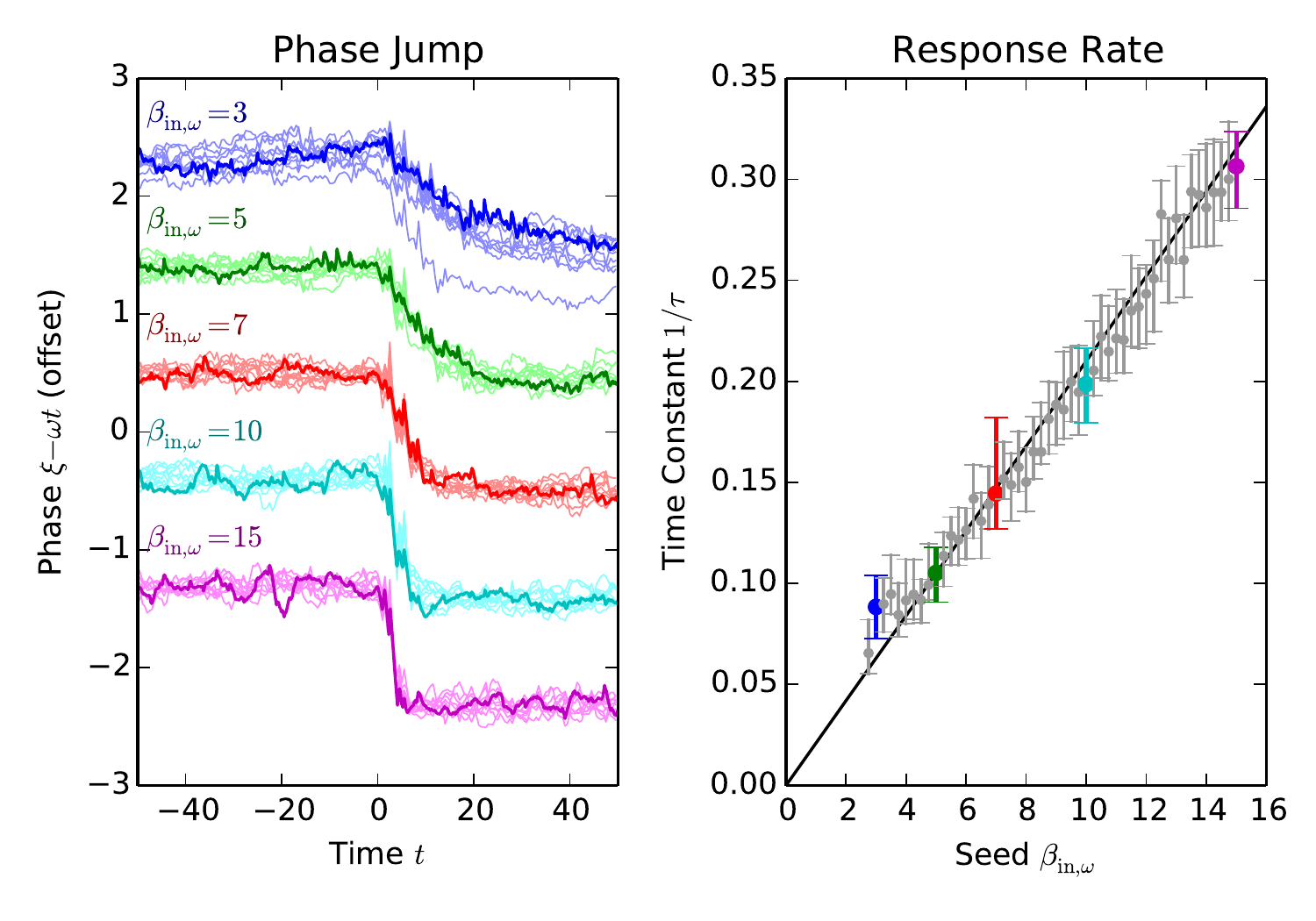}
\caption{Left: Time traces of the limit-cycle phase $\xi$ for a driven system where the seed phase jumps by one radian at $t = 0$.  Right: Response rate $1/\tau$, obtained by exponential fitting, as a function of seed amplitude $a_\omega$.  Parameters: $\Delta = -1.0, a_{\rm in} = 100$.}
\label{fig:07-f11}
\end{center}
\end{figure}

In Figure \ref{fig:07-f11}, the same free-carrier system is simulated with a seed field $\omegaIn = \omegaC = 2.27$.  However, at time $t = 0$, the phase of the input shifts by 1 radian.  For seed amplitudes $\bin{\omega} \gtrsim 3$, the system quickly realigns to the new phase, with a time-constant given by (\ref{eq:07-tau}).  From this, we can estimate $B \approx 0.02$.  This agrees with the entrainment-cone estimate to within a factor of 2; the lack of exact agreement is due to the circular cycle assumption that underlies (\ref{eq:07-lczeta}, \ref{eq:07-tau}).

\section{Applications}

\subsection{Ising Machine}
\label{sec:ising}

Many optimization problems can be recast as Ising problems, which involve finding the minimum of the Ising Hamiltonian: $H = \sum_{ij} J_{ij} \vec{\sigma}_i \cdot \vec{\sigma}_j$.  If $\sigma$ is constrained to lie on the $xy$-axis the problem is called an XY model, the each spin maps onto an angle $\sigma_i = (\cos\zeta_i, \sin\zeta_i)$ and the Hamiltonian becomes:

\beq
	U[\zeta] = \sum_{ij} J_{ij} \cos(\zeta_i - \zeta_j) \label{eq:ising-h}
\eeq

The general Ising problem for arbitrary $J_{ij}$ is NP-hard \cite{Barahona1982}.

Ising problems map naturally onto oscillator networks.  Let each Ising spin be mapped onto an oscillating free-carrier cavity.  Let each oscillator have multiple independent input and output ports.  This can be accomplished using the ``railroad topology'' of Figure \ref{fig:ising-f3}.  Suppose that an output of cavity $j$ is fed into an input of cavity $i$.  Assuming all cavities have the same limit-cycle frequency, under the assumptions of Section \ref{sec:driving}, the phase of cavity $i$ evolves as:

\begin{figure}[tbp]
\begin{center}
\includegraphics[width=1.00\columnwidth]{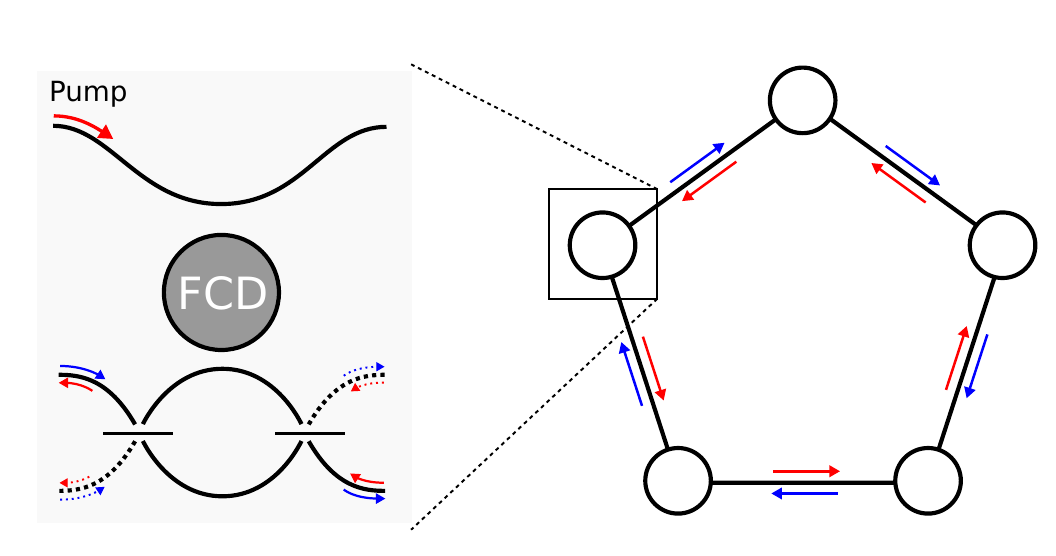}
\caption{Optical free-carrier cavity used as a node in an Ising machine.}
\label{fig:ising-f3}
\end{center}
\end{figure}

\beq
	d\zeta_i = -J_{ij} \sin(\zeta_i - \zeta_j) \label{eq:dzeta}
\eeq
where $J_{ij}$ depends on the waveguide coupling, the phase of the connection, and the limit-cycle amplitude.  It is not difficult to see that, with the appropriate connections, one can realize a cavity network that minimizes (\ref{eq:ising-h}) by the steepest-descent method.

\begin{figure}[tbp]
\begin{center}
\includegraphics[width=1.00\columnwidth]{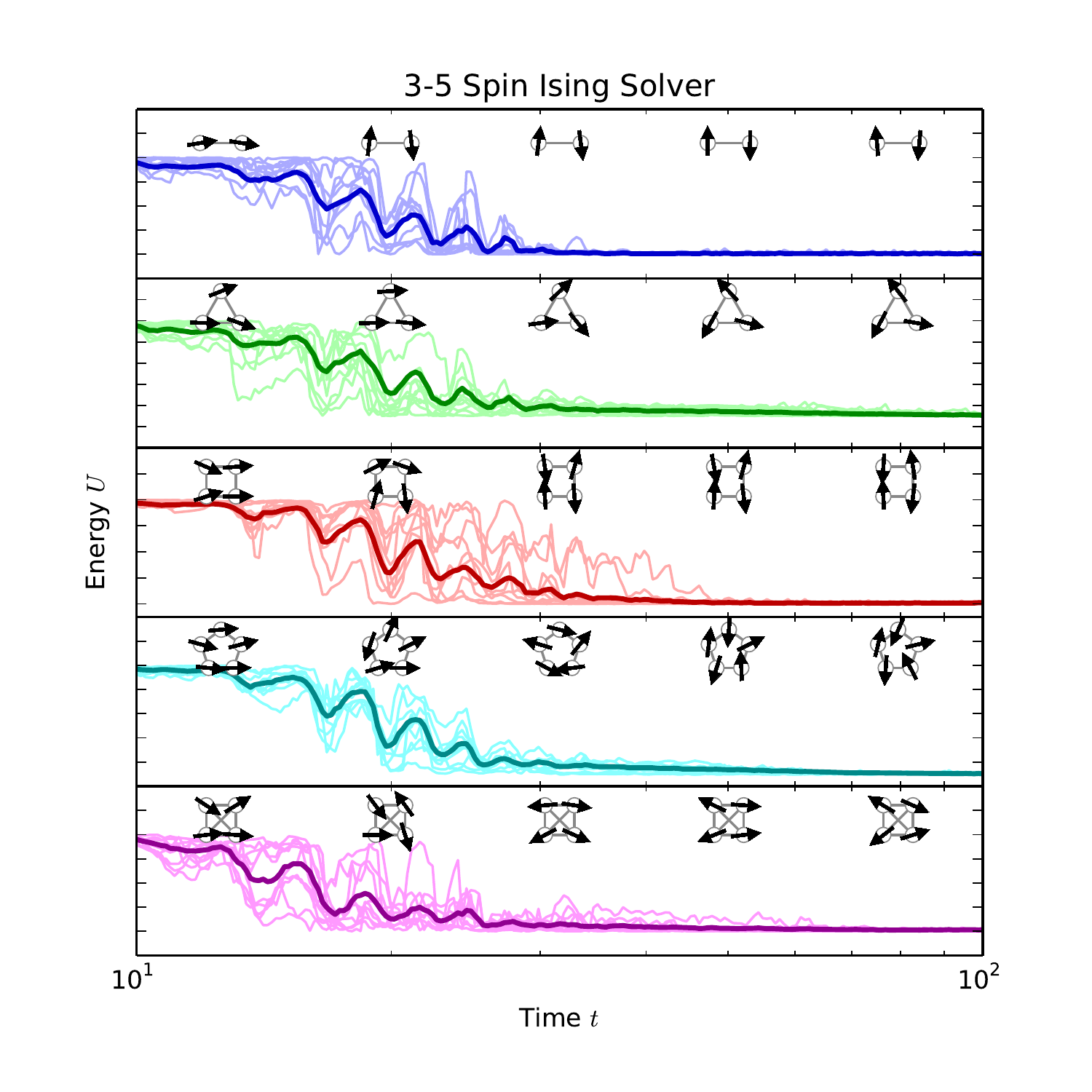}
\caption{Ising machine performance for small graphs.  Top to bottom: pair, triangle, square, pentagon, and tetrahedron.}
\label{fig:ising-f2}
\end{center}
\end{figure}

\begin{figure}[tbp]
\begin{center}
\includegraphics[width=1.00\columnwidth]{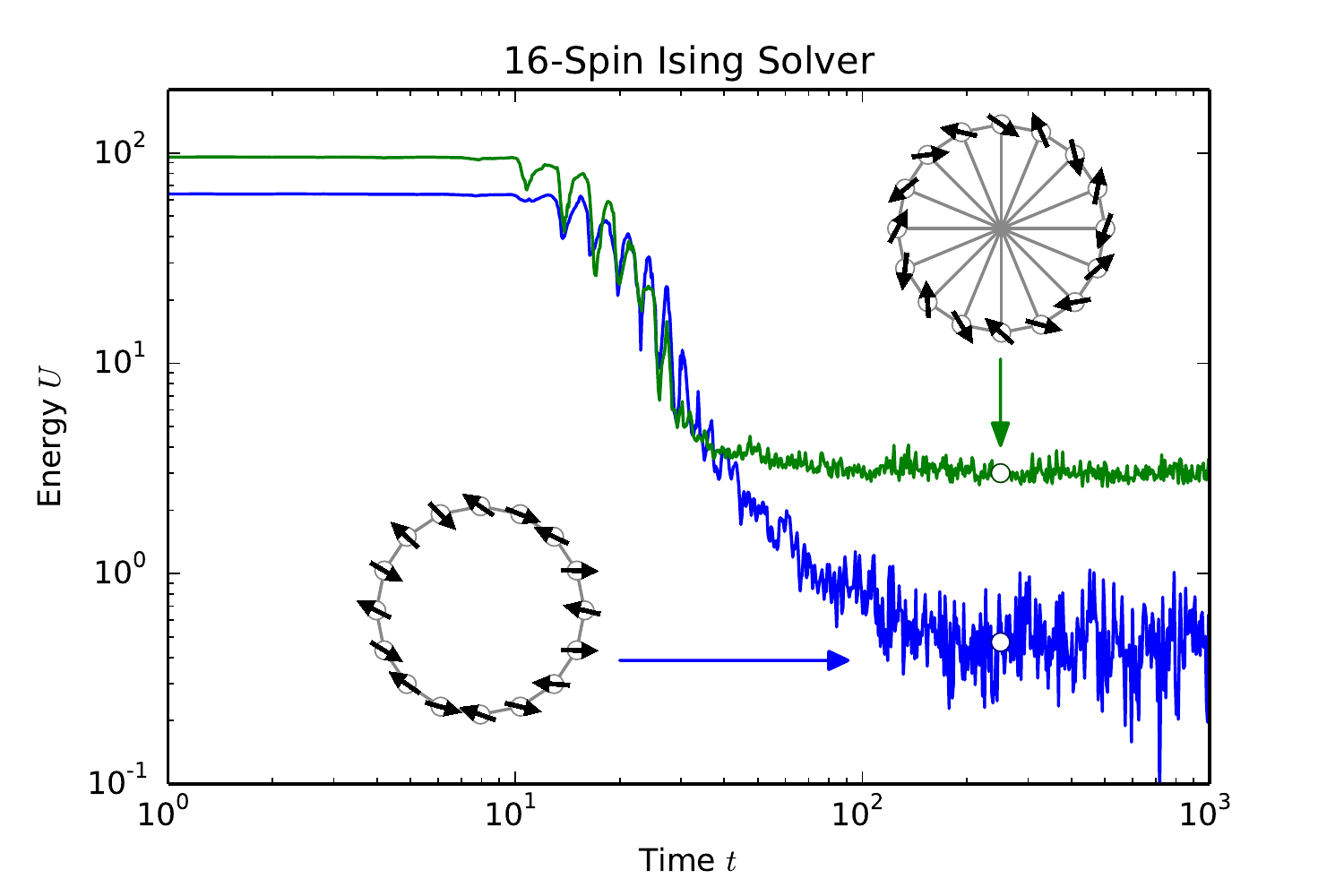}
\caption{Ising machine performance for 16-gon and frustrated 16-gon with cross-couplings.}
\label{fig:ising-f2}
\end{center}
\end{figure}

A full discussion of optical Ising machines is beyond the scope of this paper.  The concept was proposed by Utsunomiya et al. \cite{Utsunomiya2011}, who suggested implementing it using injection-locked lasers.  Recent theoretical work \cite{Wang2013} and experiments with 4-bit \cite{Marandi2014} and 16-bit \cite{KentaThesis} Ising machines using a time-multiplexed pulsed OPO show that the device matches or surpasses classical algorithms in accuracy.  However, free-carrier oscillations may be a preferable platform for Ising machines because of their low power requirements and compatibility with existing fabrication processes.

Figure \ref{fig:ising-f2} shows the simulated Ising-machine performance for antiferromagnetic couplings on five graphs: pair, triangle, square, pentagon and tetrahedron.  Of these, the pair and square have zero-energy configurations, while the rest are frustrated systems.  The square and tetrahedron were studied with an OPO Ising machine in \cite{Marandi2014}.

Larger networks also show convergence in reasonable time.  In Figure \ref{fig:ising-f2}, we plot the performance of a 16-spin network, both with a nearest-neighbor interaction and with a cross-interaction (which shows frustration).  These are the graphs studied in the OPO network of \cite{KentaThesis}.  As long as it does not get trapped in local minima, the device converges to the minimum of $U[\zeta]$ in $50-100$ cavity lifetimes.

Because the free-carrier Ising machine maps the optimization directly onto the hardware dynamics, it can achieve a per-watt performance orders or magnitude greater than a microprocessor solving the same problem.  For the network used in Figure \ref{fig:ising-f2} (see Sec.~\ref{sec:sims} for cavity parameters), during oscillation each cavity consumes $\sim 2000$ photons, or about 0.5 fJ, per cavity lifetime and takes $\sim100$ lifetimes to converge, an energy cost of $\sim 50$ fJ per spin and a computation time of $\sim 300$ ps.  A microprocessor using steepest-descent or stimulated annealing will also take $\sim 100$ steps to converge, but be required to compute (\ref{eq:dzeta}) at each step.  Since (\ref{eq:dzeta}) involves computing a trigonometric function, it will take $\sim 50$ flops and $\sim 100$ clock cycles per step \cite{AgnerFogNote}, or $\sim 5000$ flops per spin overall.  Presently, the most energy-efficient supercomputer is the L-CSC at GSI, Darmstadt, which runs at 3 GHz and requires $0.2$ nJ per flop \cite{Green500}, giving a simulation time of $\sim 3$ $\mu$s and energy cost of $\sim 1$ $\mu$J per spin.  On the basis of this rough calculation, the free-carrier Ising machine should perform $\sim 10^4\times$ faster and consume $\sim 10^7\times$ less energy.

\subsection{Free-Carrier Relay}
\label{sec:relay}

In a previous sections, we showed that free-carrier cavities can undergo spontaneous self-oscillation if driven hard enough.  Here we show that this can be used to construct a free-carrier ``relay''.  Such a device has many logic applications, including message passing algorithms for error correction \cite{Pavlichin2014}.  A relay acts like a classical CNOT gate: if the digital inputs $A, B \in \{-1, 1\}$, then the relay maps these to:

\beq
    (A,\ B) \stackrel{\rm Relay}{\longrightarrow} (A,\ A B)
\eeq
That is, output $B$ is flipped if $A = -1$.

\begin{figure}[tbp]
\begin{center}
\includegraphics[width=1.0\columnwidth]{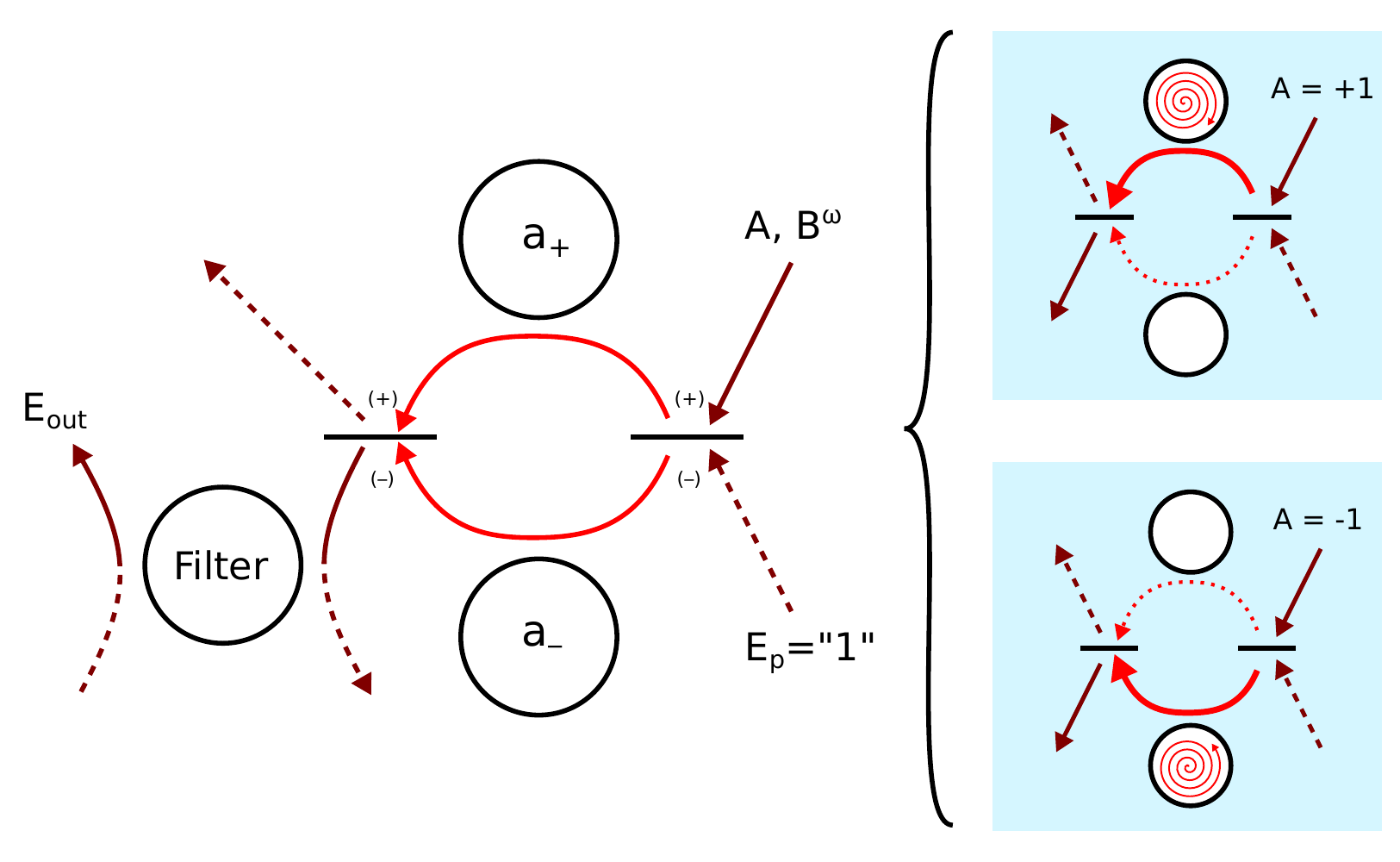}
\caption{Left: Layout of the free-carrier relay.  Right: Relay behavior when control bit $A$ is set to $+1$ (left) or $-1$ (right).}
\label{fig:07-f12}
\end{center}
\end{figure}

The relay is a circuit with two free-carrier cavities, arranged as in Figure~\ref{fig:07-f12}.  The inputs $A$ and $B$ arrive on the same channel, but are offset in frequency.  Data is encoded on the {\it phase} of the inputs ($0$ or $\pi$), not the amplitude; thus, for a fixed field amplitude $|A|$, a $1$ corresponds to $+|A|$, while $-1$ corresponds to $-|A|$.

The input is mixed with a pump field on a beamsplitter, so that the field entering cavity $a_\pm$ is:

\beq
    \bin{\pm} = \frac{A \pm E_p}{\sqrt{2}} + \frac{B e^{-i\omega t}}{\sqrt{2}}
\eeq

A free-carrier cavity will self-oscillate if the input field is stronger than some threshold: $|\binA| > \beta_{\rm th}$.  Let:

\beq
    |A| - |E_p| < \beta_{\rm th} < |A| + |E_p|.
\eeq

If $A = +1$, then the top resonator is above threshold and self-oscillates at $\omega$, while the bottom resonator does not self-oscillate.  For the $B$ field at this frequency, this means that the top channel has more gain than the bottom channel.  When these are interfered on a beamsplitter, the output at this frequency is $\frac{1}{2}(G_{\rm high} - G_{\rm low}) B e^{-i\omega t}$.  Since $G_{\rm high} > G_{\rm low}$, the phase of $B$ does not change.

On the other hand, if $A = -1$, the lower channel has higher gain.  When recombined on the beamsplitter, the output is $-\frac{1}{2}(G_{\rm high} - G_{\rm low}) B e^{-i\omega t}$ -- the phase of $B$ does flip.  This is shown in Figure \ref{fig:07-f12}.  Thus, the relay realizes the CNOT map $(A, B) \rightarrow (A, AB)$.

\begin{figure}[tbp]
\begin{center}
\includegraphics[width=1.00\columnwidth]{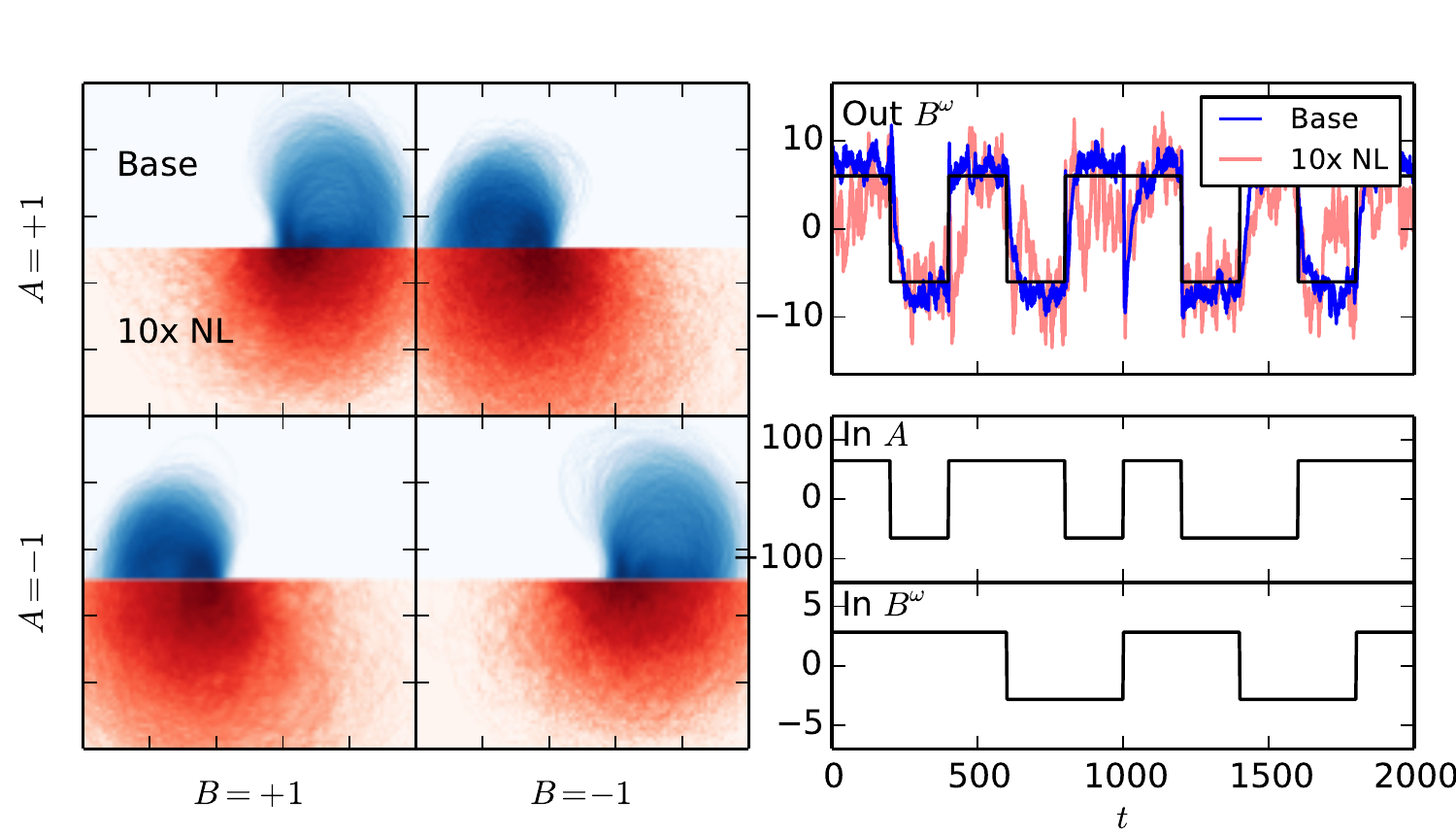}
\caption{Left: Plots of the real and imaginary parts of the rotating-frame output $B^\omega$, as a function of the input $A$ and $B^\omega$.  Right: Time trace of the relay output (top), where the inputs $A$ and $B^\omega$ are switched regularly (bottom).  Both the base (blue) and 10x NL (red) scenarios are shown.  Outputs are scaled by $\sqrt{10}$ for the 10x NL case.}
\label{fig:07-f13}
\end{center}
\end{figure}

Figure \ref{fig:07-f13} demonstrates the relay operation.  Two results are plotted: a ``base'' case with the same cavity parameters used elsewhere in the paper (blue in figure) and a hypothetical ``10x NL'' case where the nonlinearity (parameters $\delta, \beta$) has been increased by a factor of ten.  Both cavities have a detuning $\Delta = -2.0$.  In order to control the phase of the beam at $\omega$, the input $A$ must be fairly large ($A = \pm 65$ was used here, scaled by $\sqrt{10}$ for the 10x NL case).  However, the input $B$ at $\omega$ can be quite small; in the simulation taking a value of about 3.  Since the output amplitude is around 7, this provides an XOR with enough gain for a fanout of 4-5.

Both relays display the same overall behavior, but because the cavity in the 10x NL relay has a stronger nonlinearity, it operates at a lower photon number and thus the photon shot noise is more significant.  This degrades the performance of the XOR gate.  Ultimately, there is tradeoff between gate fidelity and energy consumption for free-carrier based systems.  Since this tradeoff arises from quantum mechanics, it cannot be avoided by choosing different materials or cavity designs.  The benefit of our SDE approach (\ref{eq:07-eom1}-\ref{eq:07-eom2}) is that it reveals not only the classical behavior of the relay, but also this basic quantum limit to its performance.

\section{Conclusion}

Systems with a Hopf bifurcation can perform a wide range of useful tasks with applications in sensing and photonic logic.  In this paper, we have studied the supercritical Hopf bifurcation in a semiconductor optical cavity where the dominant optical nonlinearity is due to free carrier dispersion.  Following our previous paper, we simulated the dynamics of a the free-carrier cavity using Wigner SDE's that capture both the semiclassical motion and the quantum fluctuations in photon and carrier number.

Below the bifurcation, the free-carrier optical cavity acts as a phase-insensitive amplifier.  This device is the basis for heterodyne detection, where both quadratures of the field are simultaneously measured with an added noise penalty.  The Caves bound places a lower limit on the noise, and this limit is satisfied in the non-degenerate OPO.  By contrast, the free-carrier cavity has $\sim 5\times$ more noise in the output, an effect we attribute to the incoherent nature of carrier excitation and decay.

Above the bifurcation, the device has a limit cycle.  Quantum fluctuations cause the phase of this cycle to diffuse, and the diffusion rate can be computed by linearizing the SDE's in a normal coordinate frame centered on the limit cycle.  In this limit, one can use the device to store a continuous number in the range $[0, 2\pi)$, or alternately, to perform a homodyne measurement on signals at the limit-cycle frequency.  Limits on the efficiency of homodyne measurement lead to a quantum lower bound on the limit-cycle diffusion rate.  This bound is saturated by the non-degenerate OPO, while the diffusion rate of the free-carrier cavity is $\sim 10\times$ larger.  Again, this is due to the incoherent carrier excitation and decay processes.

Limit-cycle systems are useful in logic and computing because they can be locked to external signals, and their outputs can in turn be used to lock other limit cycles.  While an analysis such large-scale networks is beyond the scope of this paper, we have explored the basic phenomenon that underlies this: entrainment in an external field.  Utilizing entrainment, we showed that the free-carrier cavity can be used to construct a coherent Ising machine that finds the minimum of a preprogrammed cost function.  With reasonable cavity parameters, such a coherent Ising machine could run $\sim 10^4\times$ faster with $\sim 10^7\times$ less energy than a comparable algorithm on a supercomputer.  In addition, we showed that entrainment can be used to construct a limit-cycle ``relay'' -- an all-optical classical CNOT gate, which has applications in message-passing schemes.

Although the free-carrier cavity is noisier and performs more poorly than quantum-limited systems like the non-degenerate OPO, it is much more convenient to build.  Free-carrier optical cavities can be built from silicon or III-V materials, which have mature and scalable fabrication processes.  In addition, the per-photon effect is much stronger, enabling operation at lower powers.  When it comes to building an actual device, these practical concerns may prevail over the theoretical elegance of quantum-limited systems.

\begin{acknowledgements}

The authors would like to thank Nikolas Tezak, Charles Santori, Jason Pelc, and Jeff Hill for helpful discussions.  This work has been supported by DARPA-MTO under award N66001-11-1-4106.  R.H.\ is supported by a Stanford Graduate Fellowship.

\end{acknowledgements}

\appendix

\section{Limit Cycles and $k$-dimensional Attractors}
\label{sec:04b-lc}

\begin{figure}[b!]
\begin{center}
\includegraphics[width=1.00\columnwidth]{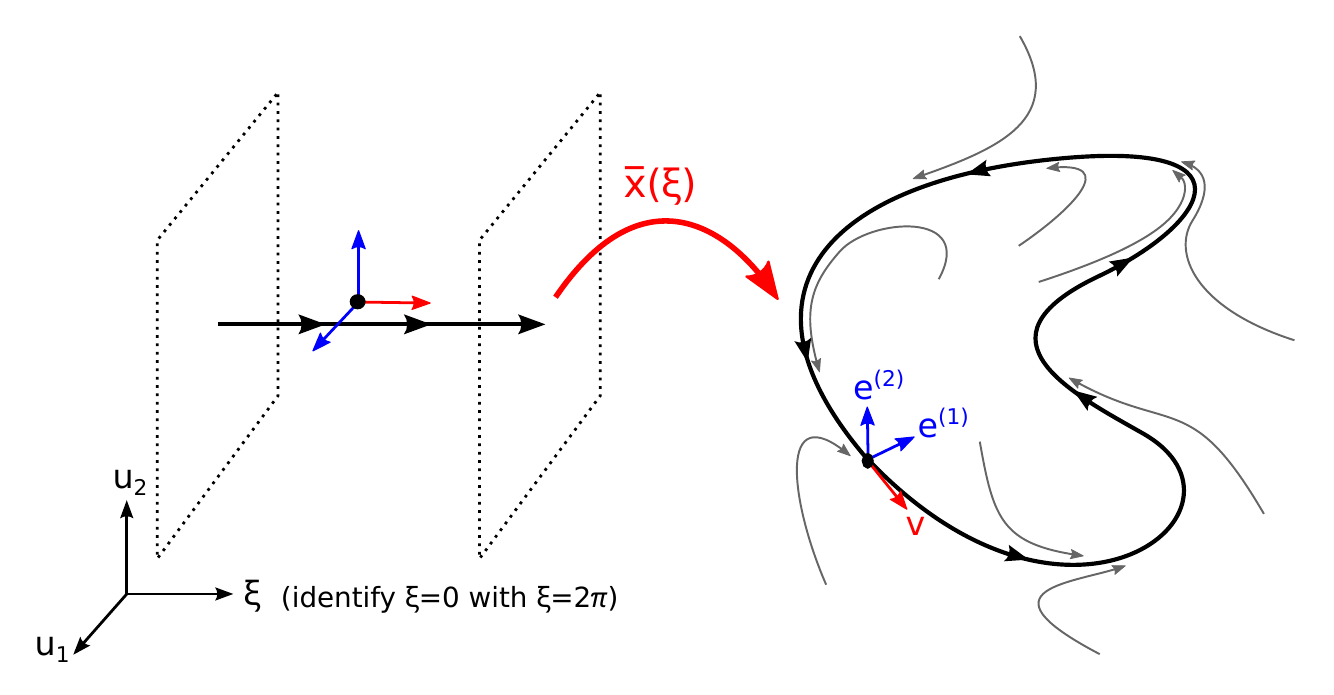}
\caption{Diagram of a limit cycle in a normal coordinate frame (left) and in the actual phase space (right), along with the transverse $e^{(i)}$ (blue) and longitudinal $\nabla_i\bar{x}$ (red) vectors.}
\label{fig:04b-f7}
\end{center}
\end{figure}

Many dynamical systems do not have a fixed point.  Instead, they have a stable limit cycle, or more generally, a stable $k$-dimensional attractor.  The $k = 1$ case corresponds to a limit cycle.  The cycle may be parameterized as follows:

\beq
	x(t) = \bar{x}(\omega t)
\eeq

where $\omega$ is the oscillation frequency.  The map $\bar{x}: \mathbb{R} \rightarrow \mathbb{R}^n$ defines the attractor's manifold, and is sufficient if we are only interested in how the system behaves without forcing.  However, the map tells us nothing about forcing or deviations from the attractor.  When noise and forcing are present, these perturbations become relevant, and we need more information about the system to handle them.

\subsection{Linearization About Attractor}

Consider a nonlinear system of differential equations of the most general form:

\beq
    dx_i = \left[f_i(x) + F_i(x, t)\right]dt + g_{ij}(x)dw_j(t) \label{eq:07-lc}
\eeq

Here, $x$ is the state of the system, $f_i(x)$ is its natural (unforced) derivative, $g_{ij}$ is the noise coupling (to Wiener process $dw_j$) and $F_i(x,t)$ is the external forcing.  In the absence of forcing, let's suppose that Equation~(\ref{eq:07-lc}) gives rise to a stable attractor $\bar{x}(\omega t)$.  This has natural period $T = 2\pi/\omega$, so $\bar{x}(\xi + 2n\pi) = \bar{x}(\xi)$ for integers $n$.  Deviations from this cycle are given by: $x(t) = \bar{x}(\omega t) + \delta x(t)$.  In the absence of noise or external forcing, the perturbations evolve as follows:

\beq
    d(\delta x_i) = \frac{\partial f_i}{\partial x_j} \delta x_i \equiv A_{ij}(\bar{x}_\xi) \delta x_j
\eeq
where $A_{ij}(x) = \partial f_i/\partial x_j$ is the Jacobian of the dynamical system; see (\ref{eq:07-linabcd}), and $\xi$ is the attractor phase, with $\bar{x}_\xi \equiv \bar{x}(\xi)$.

The key trick is to perform a coordinate transformation that separates the $dx$, and $n$-dimensional vector, into $1$ longitudinal perturbation and $n-1$ transverse perturbations.  The longitudinal perturbation keeps the system on the limit cycle, and therefore does not decay.  The transverse perturbations deviate from the limit cycle, and decay to zero as $t \rightarrow \infty$.  We denote these by $v_\xi$ and $e^{(i)}_\xi$, as follows:

\beq
	\delta x(t) = \delta\xi(t) v_\xi + \sum_{i=0}^{n-1} u_i(t) e^{(i)}_\xi \label{eq:07-lcparam}
\eeq
Here we have traded an $n$-dimensional state vector $x(t)$ for $n-1$ transverse variables $u_i(t)$ and one longitudinal variable $\delta\xi_i$.

Applying (\ref{eq:07-lcparam}) to the equations of motion with noise and forcing, we obtain:

\begin{align}
	& d(\delta\xi) v_\xi + du_i e^{(i)}_\xi = \Bigl(\underbrace{\left[A(\bar{x}_\xi) v_\xi - \omega\frac{dv_\xi}{d\xi}\right]}_{D_t v_\xi = 0} \delta\xi \nonumber \\
	& \quad + \underbrace{\left[A(\bar{x}_\xi) e^{(i)}_\xi - \omega\frac{de^{(i)}_\xi}{d\xi}\right]}_{D_t e^{(i)}_\xi} u_i\Bigr)dt + \left(F(\bar{x}_\xi, t)dt + g(\bar{x}_\xi) dw\right) \label{eq:07-lcpt}
\end{align}
(implicit summation over $i$)

The covariant derivative $D_t$ of a $\xi$-dependent vector is defined as

\beq
	D_t q_\xi \equiv A(\bar{x}_\xi)q_\xi - \omega \frac{d q_\xi}{d\xi}
\eeq

This derivative accounts for both the equations of motion and our parameterization near the limit cycle.  It is similar to the covariant derivative in Riemannian geometry \cite{WaldBook}.  Because the tangent vector $v_\xi$ always transforms into itself when propagated around the manifold, its covariant derivative is zero.  Likewise, because the transverse vectors always decay to zero, they cannot evolve into $v_\xi$; thus $D_t e_\xi^{(i)}$ has no $v_\xi$ component.

In matrix form, Equation (\ref{eq:07-lcpt}) is:

\bea
	\begin{bmatrix} v_\xi & e_\xi \end{bmatrix} \begin{bmatrix} d(\delta\xi(t)) \\ du(t) \end{bmatrix} & = &
	\begin{bmatrix} 0 & D_t e_\xi \end{bmatrix} \begin{bmatrix} \delta\xi(t) \\ u(t) \end{bmatrix}dt \nonumber \\
	& & + F(\bar{x}_\xi,t)dt + g(\bar{x}_\xi)dw
\eea

This becomes a matrix ODE:

\bea
	d\begin{bmatrix} \delta\xi(t) \\ u(t) \end{bmatrix} & = & \left(\begin{bmatrix} v_\xi & e_\xi \end{bmatrix}^{-1} \begin{bmatrix} 0 & D_t e_\xi \end{bmatrix}\right)\begin{bmatrix} \delta\xi(t) \\ u(t) \end{bmatrix} dt \nonumber \\
	& & + \begin{bmatrix} v_\xi & e_\xi \end{bmatrix}^{-1} \left(F(\bar{x}_\xi,t)dt + g(\bar{x}_\xi)dw\right) \label{eq:07-lcmat} \\
	& \equiv & \begin{bmatrix} 0 & 0 \\ 0 & A_{T} \end{bmatrix} \begin{bmatrix} \delta\xi(t) \\ u(t) \end{bmatrix}
	+ \begin{bmatrix} B_L \\ B_T \end{bmatrix} \left(F(\bar{x}_\xi,t)dt + g(\bar{x}_\xi)dw\right) \nonumber \\ \label{eq:07-lcmat2}
\eea

In the equations above, $\xi(t) = \omega t$ has a fixed time-dependence.  The dynamical variable $\delta\xi(t)$ adds a perturbation to this $\xi$.  We can roll $\delta\xi$ into $\xi$, {\it turning $\xi$ into a dynamical variable}, so the state vector becomes:

\beq
    x(t) = \bar{x}(\xi(t)) + \sum_{j=0}^{n-1}{u_j(t) e^{(j)}_{\xi(t)}}
\eeq

The matrix ODE becomes:

\begin{align}
	d\xi(t) & = \omega + B_L(\xi) \left(F(\bar{x}_\xi,t)dt + g(\bar{x}_\xi)dw\right) \label{eq:07-lc-ode1} \\
	du(t) & = A_T(\xi)u(t)dt + B_T(\xi) \left(F(\bar{x}_\xi,t)dt + g(\bar{x}_\xi)dw\right) \nonumber \\
	\label{eq:07-lc-ode2}
\end{align}

This equation captures our intuition regarding limit cycles and attractors.  External forces ($F$, $g$) can give rise to two kinds of perturbations: longitudinal (encoded in changes to $\xi$) and transverse ($u$).  Because of our choice of coordinates, {\it the perturbations evolve independently}.  The $A_T$ matrix causes transverse perturbations to decay as $t \rightarrow \infty$, while longitudinal perturbations do not.  Often, we are only interested in the longitudinal perturbations; in this case we can ignore the $u(t)$ altogether.

Altogether, we can arrive at (\ref{eq:07-lc-ode1}-\ref{eq:07-lc-ode2}) for an arbitrary limit cycle by following these four steps:

\begin{enumerate}
	\item Get equations of motion $dx = \left[f(x) + F(x, t)\right]dt + g(x)dw$
	\item Get limit cycle $\bar{x}(\xi)$ and the tangent vector $v_\xi$
	\item Find a set of vectors $e^{(i)}_\xi$ at each point $\xi$ that satisfy the following:
    	\begin{enumerate}
    		\item $\{e^{(i)}_\xi, v_\xi\}$ spans the whole vector space $\mathbb{R}^n$
    		\item Perturbations along the $\delta x \sim e^{(i)}$ eventually go to zero as $t \rightarrow \infty$
		\end{enumerate}
	\item Compute $A_T, B_L, B_T$ in Eqs.~(\ref{eq:07-lcmat}-\ref{eq:07-lcmat2})
\end{enumerate}

\subsection{Non-degenerate OPO}

Now we apply this to the non-degenerate OPO introduced in Section \ref{sec:ndopo}.  The equations of motion are reproduced below:

\bea
    d\alpha_\pm & = & \left[(-i\Delta -\kappa/2 \pm \epsilon) \alpha_\pm - \frac{\beta}{2} \left(\alpha_\pm^*\alpha_\pm \alpha_\pm - \alpha_\pm^* \alpha_\mp \alpha_\mp\right)\right]dt \nonumber \\
    & & - \sqrt{\kappa} d\bin{\pm} \mp \frac{1}{2}\sqrt{\beta}\left(\alpha_\pm dw_1 - i \alpha_\mp dw_2\right)
\eea
The limit cycle occurs at:

\beq
	|\alpha_+| = \sqrt{\frac{\epsilon - \kappa/2}{\beta/2}}
\eeq

Following the procedure above, we first find a mapping from $[0, 2\pi]$ to the limit cycle.  This is easy: $\alpha_+(\xi) = |\alpha_+| e^{-i\xi}, \alpha_-(\xi) = 0$.  Next, one needs the $v_\xi$ and $e^{(i)}$.  In terms of the basis $(\alpha_+, \alpha_-)$, a good choice is:

\beq
	\nabla_1 \bar{x}_\xi = \begin{bmatrix} -i\alpha_+ \\ 0 \end{bmatrix},\ \ \
	e^{(1)} = \begin{bmatrix} \alpha_+ \\ 0 \end{bmatrix},\ \ \
	e^{(2)} = \begin{bmatrix} 0 \\ 1 \end{bmatrix},\ \ \
	e^{(3)} = \begin{bmatrix} 0 \\ i \end{bmatrix}
\eeq

One can check that these are linearly independent (in doubled-up space) and span the whole space.  Plus, due to the symmetry of the problem, it should be pretty clear that perturbations orthogonal to the limit cycle ($e^{(1)}$) or perturbations to the $\alpha_-$ mode ($e^{(2)}, e^{(3)}$) always decay to zero.

In this case we are not concerned about deviations from the limit cycle, so there is no need to calculate the $A_T$ (which depends on covariant derivatives $D_e e_\xi$).  All we need to find is $B_L$.  At the end of the day we get the following equation of motion:

\beq
	\dot{\xi} = \Delta + {\rm Re}\,\left[\frac{-i\sqrt{\kappa}}{\alpha_+}\bin{+}\right]
\eeq

If the inputs $\bin{1}, \bin{2}$ are vacuum noise, the noise term on the right becomes

\beq
	d\xi = \Delta\,dt + \sqrt{\frac{\beta}{8} \frac{\kappa/2}{\epsilon - \kappa/2}}\,dw \label{eq:04b-ndopo-diff}
\eeq

%


\begin{thebibliography}{99}

\bibitem{Utsunomiya2011} S.~Utsunomiya, K.~Takata and Y.~Yamamoto, ``Mapping of Ising models onto injection-locked laser systems.'' \href{http://dx.doi.org/10.1364/OE.19.018091}{Optics Express {\bf 19}, 18091 (2011)}
\bibitem{Tezak2014} N.~Tezak and H.~Mabuchi, ``A coherent perceptron for all-optical learning.'' \href{http://dx.doi.org/10.1140/epjqt/s40507-015-0023-3}{EPJ Quantum Tech.\ {\bf 2}:10 (2015)}
\bibitem{Kwon2013} Y.~D.~Kwon, M.~Armen and H.~Mabuchi, ``Femtojoule-Scale All-Optical Latching and Modulation via Cavity Nonlinear Optics.'' \href{http://dx.doi.org/10.1103/PhysRevLett.111.203002}{Phys.\ Rev.\ Lett.\ {\bf 111}, 203002 (2013)}
\bibitem{KwonThesis} Y.~D.~Kwon, ``Cavity nonlinear optics with a cold atom ensemble on an atom chip: all-optical latching, modulation, and amplification.'' PhD Thesis, Stanford University (2013)
\bibitem{HiroseBook} A.~Hirose, {\it Complex-Valued Neural Networks: Theories and Applications}.  World Scientific Publishing, 2003.
\bibitem{StrogatzBook} S.~Strogatz, {\it Nonlinear Dynamics and Chaos}.  Westview Press, 1994.
\bibitem{Yamamoto1990} Y.~Yamamoto, S.~Machida, S.~Saito, N.~Imoto et al., ``Quantum Mechanical Limit in Optical Precision Measurement and Communication.'' \href{http://dx.doi.org/10.1016/s0079-6638(08)70289-0}{Prog. Optics  {\bf 28}, 87 (1990)}
\bibitem{Reid1987} M.~D.~Reid and P.~D.~Drummond, ``Quantum Correlations of Phase in Nondegenerate Parametric Oscillation.'' \href{http://dx.doi.org/10.1103/physrevlett.60.2731}{Phys.\ Rev.\ Lett.\ {\bf 60}, 2731 (1988)}
\bibitem{Armen2006} M.~A.~Armen and H.~Mabuchi, ``Low-lying bifurcations in cavity quantum electrodynamics.'' \href{http://dx.doi.org/10.1103/PhysRevA.73.063801}{Phys.\ Rev.\ A {\bf 73}, 063801 (2006)}
\bibitem{Poberaj2012} G.~Poberaj, H.~Hu, W.~Sohler and P.~Gunter, ``Lithium niobate on insulator (LNOI) for micro-photonic devices.'' \href{http://dx.doi.org/10.1002/lpor.201100035}{Laser and Photonics Reviews {\bf 6}, 488 (2012)}
\bibitem{Notomi2010} M.~Notomi, ``Manipulating light with strongly modulated photonic crystals.'' \href{http://dx.doi.org/10.1088/0034-4885/73/9/096501}{Rep.\ Prog.\ Phys.\ {\bf 73}, 096501 (2010)}
\bibitem{Malaguti2011} S.~Malaguti, G.~Bellanca, A.~de~Rossi, S.~Combrie and S.~Trillo, ``Self-pulsing driven by two-photon absorption in semiconductor nanocavities.'' \href{http://dx.doi.org/10.1103/physreva.83.051802}{Phys.\ Rev.\ A {\bf 83}, 051802(R) (2011)}
\bibitem{Malaguti2013} S.~Malaguti, G.~Bellanca and S.~Trillo, ``Low-power spontaneous oscillations driven by band-filling effect.'' \href{http://dx.doi.org/10.1364/OL.38.004366}{Optics Letters {\bf 38}, 4366 (2013)}
\bibitem{Chen2012} S.~Chen, L.~Zhang, Y.~Fei, and T.~Cao, ``Bistability and self-pulsation phenomena in silicon microring resonators based on nonlinear optical effects.'' \href{http://dx.doi.org/10.1364/oe.20.007454}{Optics Express {\bf 20}, 7454 (2012)}
\bibitem{Nozaki2010} K.~Nozaki, T.~Tanabe, A.~Shinya et al., ``Sub-femtojoule all-optical switching using a photonic-crystal nanocavity.'' \href{http://dx.doi.org/10.1038/NPHOTON.2010.89}{Nature Photonics {\bf 4}, 477 (2010)}
\bibitem{Paper1} R.~Hamerly and H.~Mabuchi, ``Quantum Noise of Free-Carrier Dispersion in Semiconductor Optical Cavities.'' \href{http://arxiv.org/abs/1504.04409}{Phys.\ Rev.\ A (to be published), arXiv:1504.04409}.
\bibitem{Caves1982} C.~Caves, ``Quantum limits on noise in linear amplifiers.'' \href{http://dx.doi.org/10.1103/physrevd.26.1817}{Phys.\ Rev.\ D {\bf 26}, 1817 (1982)}
\bibitem{VanVar2012} T.~Van~Vaerenbergh, M.~Fiers, J.~Dambre, and P.~Bienstman, ``Simplified description of self-pulsation and excitability by thermal and free-carrier effects in semiconductor microcavities.'' \href{http://dx.doi.org/10.1103/PhysRevA.86.063808}{Phys.\ Rev.\ A {\bf 86}, 063808 (2012)}
\bibitem{JohnsonThesis} T.~Johnson, ``Silicon Microdisk Resonators for Nonlinear Optics and Dynamics.'' Ph.D.\ thesis, Caltech, 2009.
\bibitem{Johnson06} T.~Johnson, M.~Borselli, and O.~Painter, ``Self-induced optical modulation of the transmission through a high-Q silicon microdisk resonator.'' \href{http://dx.doi.org/10.1364/OPEX.14.000817}{Optics Express {\bf 14}, 817 (2006)}
\bibitem{Gardiner1985} C.~W.~Gardiner and M.~J.~Collett, ``Input and output in damped quantum systems: Quantum stochastic differential equations and the master equation.'' \href{http://dx.doi.org/10.1103/physreva.31.3761}{Phys.\ Rev.\ A {\bf 31}, 3761 (1985)}
\bibitem{WallsMilburn} D.~F.~Walls and G.~J.~Milburn, {\it Quantum Optics}.  Springer, 1994.
\bibitem{QuantumNoise} C.~W.~Gardiner and P.~Zoller, {\it Quantum noise: a handbook of Markovian and non-Markovian quantum stochastic methods with applications to quantum optics}, 3rd ed.  Springer, 2004.
\bibitem{Santori2014} C.~Santori, J.~Pelc, R.~Beausoleil et al., ``Quantum Noise in Large-Scale Coherent Nonlinear Photonic Circuits.'' \href{http://dx.doi.org/10.1103/physrevapplied.1.054005}{Phys.\ Rev.\ Applied {\bf 1}, 054005 (2014)}
\bibitem{Gronchi1978} M.~Gronchi and L.~Lugiato, ``Fokker-planck equation for optical bistability.'' \href{http://dx.doi.org/10.1007/BF02776284}{Lett.\ Nuovo Cimento {\bf 23}, 593 (1978)}
\bibitem{Agrawal1979} G.~P.~Agrawal and H.~J.~Carmichael, ``Optical bistability through nonlinear dispersion and absorption.'' \href{http://dx.doi.org/10.1103/PhysRevA.19.2074}{Phys.\ Rev.\ A {\bf 19}, 2074 (1979)}
\bibitem{Wiesenfeld1986} K.~Wiesenfeld and B.~McNamara, ``Small-signal amplification in bifurcating dynamical systems.'' \href{http://dx.doi.org/10.1103/PhysRevA.33.629}{Phys.\ Rev.\ A {\bf 33}, 629 (1986)}
\bibitem{GoughJames2008} J.~Gough and M.~James, ``The Series Product and Its Application to Quantum Feedforward and Feedback Networks.'' \href{http://dx.doi.org/10.1109/TAC.2009.2031205}{IEEE TAC {\bf 54}, 2530 (2009)}
\bibitem{Gough2009Sq} J.~E.~Gough and S.~Wildfeuer, ``Enhancement of field squeezing using coherent feedback.'' \href{http://dx.doi.org/10.1103/physreva.80.042107}{Phys.\ Rev.\ A {\bf 80}, 042107 (2009)}
\bibitem{Gough2010} J.~E.~Gough, M.~R.~James, and H.~I.~Nurdin, ``Squeezing components in linear quantum feedback networks.'' \href{http://dx.doi.org/10.1103/physreva.81.023804}{Phys.\ Rev.\ A {\bf 81}, 023804 (2010)}
\bibitem{WigginsBook} S.~Wiggins, {\it Introduction to Applied Nonlinear Dynamical Systems and Chaos}.  Springer, 1990.
\bibitem{Wolinsky1988} M.~Wolinsky and H.~J.~Carmichael, ``Quantum noise in the parametric oscillator: From squeezed states to coherent-state superpositions.'' \href{http://dx.doi.org/10.1103/PhysRevLett.60.1836}{Phys.\ Rev.\ Lett.\ {\bf 60}, 1836 (1988)}
\bibitem{Mirrahimi2013} M.~Mirrahimi, Z.~Leghtas, V.~Albert, S.~Touzard et al., ``Dynamically protected cat-qubits: a new paradigm for universal quantum computation.'' \href{http://dx.doi.org/10.1088/1367-2630/16/4/045014}{New J.\ Phys.\ {\bf 16}, 045014 (2014)}
\bibitem{Mabuchi2012} H.~Mabuchi, ``Qubit limit of cavity nonlinear optics.'' \href{http://dx.doi.org/10.1103/PhysRevA.85.015806}{Phys.\ Rev.\ A {\bf 85}, 015806 (2012)}
\bibitem{Barahona1982} F.~Barahona, ``On the computational complexity of Ising spin glass models.'' \href{http://dx.doi.org/10.1088/0305-4470/15/10/028}{J.\ Phys.\ A: Math.\ Gen.\ {\bf 15}, 3241 (1982)}
\bibitem{Wang2013} Z.~Wang, A.~Marandi, K.~Wen, R.~L.~Byer, and Y.~Yamamoto, ``Coherent Ising machine based on degenerate optical parametric oscillators.'' \href{http://dx.doi.org/10.1103/PhysRevA.88.063853}{Phys.\ Rev.\ A {\bf 88}, 063853 (2013)}
\bibitem{Marandi2014} A.~Marandi, Z.~Wang, K.~Takata, R.~L.~Byer and Y.~Yamamoto, ``Network of time-multiplexed optical parametric oscillators as a coherent Ising machine.'' \href{http://dx.doi.org/doi:10.1038/nphoton.2014.249}{Nature Photonics {\bf 8}, 937 (2014)}
\bibitem{KentaThesis} K.~Takata, Ph.D.\ thesis, University of Tokyo, 2015.
\bibitem{AgnerFogNote} A.~Fog, Technical University of Denmark.  \href{http://www.agner.org/optimize/instruction_tables.pdf}{\it Instruction Tables}.
\bibitem{Green500} \href{http://www.green500.org/lists/green201411}{Green500 List -- November 2014}
\bibitem{Pavlichin2014} D.~Pavlichin and H.~Mabuchi, ``Photonic circuits for iterative decoding of a class of low-density parity-check codes.'' \href{http://dx.doi.org/10.1088/1367-2630/16/10/105017}{New J.\ Phys.\ {\bf 16}, 105017 (2014)}
\bibitem{WaldBook} R.~Wald, {\it General Relativity}.  University of Chicago Press, 1984.



\end{thebibliography}
\end{document}